\documentclass[english,11pt,aps,prd,a4paper,preprintnumbers,floatfix,nofootinbib,showpacs,superscriptaddress, notitlepage]{revtex4-1} 

\newcommand{\AddrUNAM}{Instituto de F\'isica, Universidad Nacional Aut\'onoma de M\'exico, A.P. 20-364, Ciudad de M\'exico 01000, M\'exico.}

\usepackage{booktabs}
\usepackage{color}
\usepackage{graphicx}
\usepackage{multirow}
\usepackage{amsmath}
\usepackage{amssymb}
\usepackage[colorlinks=true,citecolor=darkred,urlcolor=darkred, pdfborder={0 0 0}]{hyperref}
\usepackage[normalem]{ulem}
\definecolor{darkred}{rgb}{0.6,0,0}
\definecolor{drkgrn}{RGB}{0, 51, 0}
\definecolor{gray}{RGB}{128, 128, 128}
\usepackage{soul}

\def\beq{\begin{equation}}
\def\eeq{\end{equation}}

\newcommand{\eps}{\varepsilon}
\def\cevns{CE$\nu$NS~}

\begin{document}

\title{Non-standard neutrino interactions in $U(1)'$ model  after COHERENT data}

\author{L.  J.  Flores}\email{luisjf89@fisica.unam.mx}\affiliation{\AddrUNAM}
\author{Newton Nath}\email{newton@fisica.unam.mx}\affiliation{\AddrUNAM}
\author{Eduardo Peinado} \email{epeinado@fisica.unam.mx}\affiliation{\AddrUNAM}

\begin{abstract}
{\noindent
We explore the potential to prove light extra gauge $Z^\prime$ boson inducing non-standard neutrino interactions (NSIs) in the  coherent-elastic neutrino-nucleus scattering (CE$ \nu $NS) experiments. We intend to examine  how the latest COHERENT-CsI and CENNS-10 data can constrain this model. A detailed investigation for the upcoming Ge, LAr-1t, and NaI detectors of COHERENT collaboration has also been made. Depending on numerous other constraints  coming from  oscillation experiments, muon $ (g-2) $, beam-dump experiments, LHCb, and reactor experiment CONUS, we explore the parameter space in $Z^\prime$ boson mass vs coupling constant plane. Moreover, we study the predictions of two-zero textures that are allowed in the concerned model in light of the latest global-fit data.

}
\end{abstract}

\maketitle

\section{Introduction}
Oscillations of neutrinos among different flavors are now a well established phenomenon from various experimental searches, which implies that the neutrinos carry non-zero masses and their different flavor are substantially mixed \cite{Tanabashi:2018oca}. 
Currently, we have fairly good understanding of all the neutrino oscillation parameters in three-flavor paradigm, except the Dirac  CP violating phase~\cite{Capozzi:2016rtj,Esteban:2016qun,deSalas:2018bym}.
This led us into an era of precision measurements in the leptonic sector, where it is possible to observe sub-leading effects originating from physics beyond the Standard Model (SM). Furthermore, this may affect the propagation of neutrinos and eventually it may impact the measurements of three-flavor neutrino oscillation parameters. 
Among various new physics  scenarios beyond the standard three-flavor neutrino oscillations, non-standard neutrino interactions (NSIs) can be induced by the new physics beyond the SM (BSM).
In literature, they are traditionally described by the dimension-6 four-fermion operators of the form \cite{Wolfenstein:1977ue},
\begin{equation}
\label{eq:NSI}
\mathcal
{L}_\text{NSI} \supset 
(\overline{\nu}_\alpha \gamma^{\rho} 
 \nu_\beta)
(\bar{f} \gamma_{\rho} f)
\epsilon^{f}_{\alpha\beta} 
+ \text{h.c.}
\end{equation}
where $\epsilon^{f}_{\alpha\beta}$ represent NSI parameters and 
$\alpha, \beta = e, \mu, \tau$, $f = e, u, d$.
The importance of NSIs were discussed well before the establishment of 
neutrino oscillation phenomena by a number of authors in~\cite{Wolfenstein:1977ue, 
Valle:1987gv, Roulet:1991sm, Guzzo:1991hi}. For a detailed model-independent review of NSIs and their phenomenological consequences see Refs.\cite{Ohlsson:2012kf, Miranda:2015dra, Farzan:2017xzy} and  the references therein.

In  past a few years there are many BSM models that have addressed NSIs. Some of the popular models where NSIs can be present are the flavor-sensitive,  $ Z^{\prime} $ mediated  $U(1)^\prime$ extended gauge models~\cite{Barranco:2005yy,Scholberg:2005qs,Denton:2018xmq,Heeck:2018nzc,Han:2019zkz}~\footnote{Note that some recent studies of NSIs considering heavy charged singlet and/or doublet scalars have been performed in \cite{Forero:2016ghr,Dey:2018yht,Liao:2019qbb}.}.
In these scenarios,  an extra gauged $U(1)$ is added to the SM gauge group, where the corresponding symmetry breaking leads to a new gauge boson  $Z^\prime$. In literature, numerous studies have been performed based on $U(1)^\prime$ model, ranging from flavor models to GUTs scenarios~\cite{Marshak:1979fm,Mohapatra:1980qe, Baek:2001kca,Khalil:2006yi}.
%
Dark matter phenomenology based on such symmetry has been addressed in~\cite{Okada:2010wd,Okada:2012sg,Basso:2012ti,Basak:2013cga}.
Furthermore, to provide strong
constraints between the mass and gauge coupling of associate gauge boson  $Z^\prime$,   a  large variety
of measurements have been performed, such as rare decays, anomalous magnetic moments
of the electron or muon, electroweak precision tests, and direct searches at the LHC
~\cite{Erler:1999ub,Langacker:2008yv,Basso:2008iv,Erler:2009jh,Salvioni:2009mt,
Salvioni:2009jp,Ekstedt:2016wyi,Bandyopadhyay:2018cwu,Aebischer:2019blw,Dudas:2013sia,Okada:2018tgy,Deppisch:2019ldi}.
However, our main focus is to examine the importance of non-standard neutrino interactions within the gauge extended framework of the SM.

At the current juncture, the latest probe of NSIs come from the observation of Coherent Elastic $\nu$-Nucleus Scattering (CE$\nu$NS) processes,  first observed by the COHERENT collaboration~\cite{Akimov:2017ade} in 2017 using cesium iodide (CsI) scintillation detector as a target.
They have reported their first detection of  $\nu$-nucleus scattering at $6.7$ $\sigma$ \cite{Akimov:2017ade}. The measurement is consistent  with the SM expectations at 1.5 $\sigma$ and within the SM, it is induced by the $Z$ boson exchange~\cite{Freedman:1973yd}.
On the other hand,  the first measurement of \cevns on argon by the COHERENT collaboration has been reported in~\cite{Akimov:2020pdx}, with more than $3~\sigma$ significance level. They have used the CENNS-10 liquid argon detector, providing the lightest nucleus measurement of \cevns.
It is important to study these processes because of their ability to probe the SM parameters at low momentum transfer~\cite{Scholberg:2005qs,Lindner:2016wff,Deniz:2017zok,Miranda:2019wdy}, new physics scenarios, like NSIs~\cite{ Liao:2017uzy,AristizabalSierra:2018eqm,Giunti:2019xpr,Esteban:2019lfo,Coloma:2019mbs,Khan:2019cvi}, 
nuclear physics parameters~\cite{Cadeddu:2017etk,Papoulias:2019lfi,Canas:2019fjw}, neutrino  electromagnetic properties~\cite{Cadeddu:2018dux,Miranda:2019wdy}, and sterile neutrinos~\cite{Miranda:2019skf,Blanco:2019vyp,Berryman:2019nvr}. 
Recently, it has been addressed in Refs.~\cite{Dutta:2015vwa,Dent:2016wcr,Billard:2018jnl,Datta:2018xty,Denton:2018xmq,Farzan:2018gtr,AristizabalSierra:2019ufd,Dutta:2019eml,AristizabalSierra:2019ykk,Brdar:2018qqj} that light mediators may be accessible to \cevns experiments. Considering a new gauge boson  $Z^\prime$ associated with new $U(1)^\prime$ symmetry from \cevns have been studied in ~\cite{Liao:2017uzy,Kosmas:2017tsq,Denton:2018xmq,Heeck:2018nzc,Han:2019zkz,Miranda:2020zji,Papoulias:2019xaw}. 

In this work, we investigate non-standard neutrino interactions arising from a new gauge boson $ Z^{\prime} $ associated with an extra $U(1) ^\prime=U(1)_{B-2L_\alpha-L_\beta}$ symmetry, where $\alpha \neq \beta=e, \mu, \tau$.
Considering the combined effect of different experimental constraints coming from the COHERENT collaboration, oscillation data,  beam-dump experiments, and the LHCb dark photon searches,  we examine the allowed region in $ M_{Z^\prime} $ vs the coupling constant $ g^\prime $ plane that can lead to possible NSIs. In addition, we also explore the potential of reactor based CE$\nu$NS experiment like the COherent NeUtrino Scattering experiment (CONUS)~\cite{Lindner:2016wff}.
Bounds arising from other processes like   anomalous magnetic moment of muon i.e., $ (g-2)_{\mu} $, from astrophysical observations such as Big Bang Nucleosynthesis (BBN) and Cosmic Microwave Background (CMB) have also been shown for the comparison.

 Furthermore, by introducing two different  $U(1)^\prime$ breaking scalar fields, it has been observed that the neutrino oscillation parameters are in well agreement with the current global-fit data~\cite{Capozzi:2016rtj,Esteban:2016qun,deSalas:2018bym}.
We end up with four different scenarios compatible with the neutrino oscillation data, depending on  $U(1)^\prime$ charges of the model, namely $  U(1)_{B-L_\mu-2L_\tau}$, $  U(1)_{B-2L_\mu-L_\tau}$, $  U(1)_{B-L_e-2L_\tau}$, and $  U(1)_{B-L_e-2L_\mu}$\footnote{The other possible combinations cannot explain oscillation data.}. 
It has been realized that these four scenarios give rise to four different two-zero textures for the light neutrino mass matrix, namely, $A_1$, $A_2$, $B_3$ and $B_4$.

Each of these four cases have their own NSI structure. We  explore the impact for each model considering current  COHERENT~\cite{Akimov:2017ade} data and for the future CE$\nu$NS experiments. Other neutrino phenomenology, such as the predictions for the neutrino-less double beta ($ 0\nu\beta\beta $) decay and the prediction for the lightest neutrino mass have also been discussed.

The remainder of the paper is outlined as follows. In next Sec.~(\ref{sec:NSIs}), we give a brief  description of  non-standard interactions (NSIs) and their latest bounds. The theoretical set-up of the model has been discussed in Sec.~(\ref{sec:model}).
Sec.~(\ref{sec:coherent}) is devoted to CE$ \nu $NS processes as well as other experimental and numerical details. The principle results of the paper has also been discussed in this section.  Later, in  Sec.~(\ref{sec:two-zero}) phenomenology of two-zero textures have been addressed. 
We summarize our findings in Sec.~(\ref{sec:conclusion}). Appendix \ref{sec:anomalycanc} has dealt with the anomaly cancellation of the $U(1)^\prime$ symmetry and the light neutrino mass under type-I seesaw mechanism has been discussed in appendix \ref{sec:appB}.

\section{Non-standard neutrino interactions} \label{sec:NSIs}
Here we present a general description of the  non-standard interactions  involving neutrinos. We consider the effect of neutral-current NSI in  presence of matter which is describe by the dimension-6 four-fermion operators of the form \cite{Wolfenstein:1977ue},
\begin{equation}
\label{eq:NSI}
- \mathcal {L}^{NC}_\text{NSI} =   2\sqrt{2} G_F 
                       \epsilon^{fC}_{\alpha\beta} ~ (\overline{\nu}_\alpha \gamma^{\rho} 
                       P_L \nu_\beta)
                      (\bar{f} \gamma_{\rho} P_C f)                      
                      + \text{h.c.} \;,
\end{equation}
where $\epsilon^{f C}_{\alpha\beta}$ are NSI parameters, 
$\alpha, \beta = e, \mu, \tau$, $C = L,R$, denotes the chirality, $f = e, u, d$, and 
$G_{F}$ is the Fermi constant~\footnote{Here we neglect  the effect of charged-current NSIs which mainly affect 
the  production and detection of neutrinos~\cite{Khan:2013hva, Ohlsson:2013nna, Girardi:2014kca, DiIura:2014csa, 
Agarwalla:2014bsa, Blennow:2015nxa}.}.
The Hamiltonian in presence of matter NSI, in the 
flavor basis, can be written as,
\begin{equation} 
\label{eq:nsi_hamil}
H = \frac{1}{2E} \left[U
                  \text{diag}(0,\Delta m^2_{21},\Delta m^2_{31})
                  U^\dagger 
                  + {\rm diag}(A,0,0) 
                  + A \epsilon_{\alpha\beta}
                  \right]\,,
\end{equation}
where $U$ is the Pontecorvo-Maki-Nakagawa-Sakata (PMNS) mixing 
matrix~\cite{Patrignani:2016xqp}, $\Delta m^2_{ij}=m^2_i-m^2_j$  ($ i<j = 1,2,3 $), and $A\equiv 2\sqrt2 G_F N_e E$ represents the 
 potential due to the standard matter interactions  of neutrinos and $ \epsilon_{\alpha\beta}\ $  can be written as
\begin{equation} 
\label{eq:potential}
\epsilon_{\alpha\beta}\ =  \left(\begin{array}{ccc}
 \epsilon_{ee} & \epsilon_{e\mu} & \epsilon_{e\tau} 
\\
\epsilon^{*}_{e\mu}  & \epsilon_{\mu\mu} & \epsilon_{\mu\tau}
\\
\epsilon^{*}_{e\tau} & \epsilon^{*}_{\mu\tau} & \epsilon_{\tau\tau}
\end{array}\right)\, \;,
\end{equation}
where $ \epsilon_{\alpha\beta} = |\epsilon_{\alpha\beta}| e^{i\phi_{\alpha\beta}} $ for $ \alpha \neq\beta $.
In general,  the elements of $\epsilon_{\alpha\beta}$ are complex for 
$\alpha\neq \beta$, whereas diagonal elements are real due to
the Hermiticity of the Hamiltonian as given by Eq.~\eqref{eq:nsi_hamil}. For the matter NSI, 
$\epsilon_{\alpha\beta}$  can be  defined as,
\begin{equation}
 \epsilon_{\alpha\beta}  = \sum_{f,C}  \epsilon^{fC}_{\alpha\beta} \dfrac{N_f}{N_e} ~,
\end{equation}
where  $N_f$ is the number density of fermion $f$ and 
$\epsilon^{fC}_{\alpha\beta} = \epsilon^{f L}_{\alpha\beta} 
+ \epsilon^{f R}_{\alpha\beta}$. In case of the Earth matter, one can 
assume that the number densities of electrons, protons, and 
neutrons are equal (i.e. $ N_p \simeq N_n =N_e $), in such a case 
$ N_u \simeq N_d \simeq  3N_e $ and one can write,
\begin{equation}
\epsilon_{\alpha\beta} = \sqrt{\sum_C \left(
         (\epsilon^{e C}_{\alpha\beta})^{2} 
         + (3 \epsilon^{u C}_{\alpha\beta})^{2} 
         + (3\epsilon^{d C}_{\alpha\beta})^{2} 
         \right)} \;.
\label{nsiee1}
\end{equation}
In Table~\ref{tab:nsiConstraints}, we give recent constraints for NSIs  obtained from a combined analysis of oscillation experiments and COHERENT measurements~\cite{Esteban:2018ppq} at $2\sigma$ C.L.
\begin{widetext}
 \begin{center}
\begin{table}[t] 
\begin{tabular}{|cc|cc|}  \hline
    	& OSC &  & OSC + COHERENT \\ \hline \hline
    	$\eps^u_{ee}-\eps^u_{\mu\mu}$ 		& $[-0.020, 0.456]$ & \quad $\eps^u_{ee}$ & $[-0.008, 0.618]$ \\
    	$\eps^u_{\tau\tau}-\eps^u_{\mu\mu}$ & $[-0.005, 0.130]$ & \quad  $\eps^u_{\mu\mu}$ & $[-0.111, 0.402]$ \\
    	 									& & \quad  $\eps^u_{\tau\tau}$ & $[-0.110, 0.404]$ \\ \hline
    	$\eps^u_{e\mu}$ 					& $[-0.060, 0.049]$ & \quad  $\eps^u_{e\mu}$ & $[-0.060, 0.049]$ \\
    	$\eps^u_{e\tau}$ 					& $[-0.292, 0.119]$ & \quad  $\eps^u_{e\tau}$ & $[-0.248, 0.116]$ \\
    	$\eps^u_{\mu\tau}$ 					& $[-0.013, 0.010]$ & \quad  $\eps^u_{\mu\tau}$ & $[-0.012, 0.009]$ \\ \hline
    	$\eps^d_{ee}-\eps^d_{\mu\mu}$  		& $[-0.027, 0.474]$ & \quad  $\eps^d_{ee}$ & $[-0.012, 0.565]$ \\
    	$\eps^d_{\tau\tau}-\eps^d_{\mu\mu}$ & $[-0.005, 0.095]$ & \quad  $\eps^d_{\mu\mu}$ & $[-0.103, 0.361]$ \\
    	 									& & \quad  $\eps^d_{\tau\tau}$ & $[-0.102, 0.361]$ \\ \hline
    	$\eps^d_{e\mu}$ 					& $[-0.061, 0.049]$ & \quad  $\eps^d_{e\mu}$ & $[-0.058, 0.049]$ \\
    	$\eps^d_{e\tau}$ 					& $[-0.247, 0.119]$ & \quad  $\eps^d_{e\tau}$ & $[-0.206, 0.110]$ \\
    	$\eps^d_{\mu\tau}$ 					& $[-0.012, 0.009]$ & \quad  $\eps^d_{\mu\tau}$ & $[-0.011, 0.009]$\\ \bottomrule
\hline
  \end{tabular}
\caption{\footnotesize Recent constraints for the NSI parameters $\eps^u_{\alpha\beta}$ and $\eps^d_{\alpha\beta}$, at $2\sigma$ C.L., obtained from the combined analysis of  the oscillation experiments and COHERENT measurements~\cite{Esteban:2018ppq}.}
    \label{tab:nsiConstraints}
\end{table}
\end{center}
\end{widetext}
Having introduced general descriptions of NSI and its bounds, in next section we describe our model in great details.
\section{The setup}\label{sec:model}

 In this work, we  extend the SM gauge group to an anomaly free $U(1)^\prime=U(1)_{B-2L_\alpha - L_\beta}$
~\footnote{The anomaly cancelation conditions have been addressed in the appendix~\ref{sec:anomalycanc}.}, where $\alpha$ and $\beta$ can be $e,~\mu$ and $\tau$.
 In our framework,  two different lepton flavors are coupled to the new gauge interaction. The relevant charge assignments for the lepton fields as well as the scalar fields that trigger the $U(1)^\prime$ gauge symmetry breaking are listed in Table~\ref{tab:charges}.
 In this prescription, we include two scalar fields $\phi_1$ and $\phi_2$ transforming as $1$ and $2$ under $U(1)^\prime$,  respectively. It is worth to mention that this $U(1)^\prime$ interaction is not flavor violating. These scalar fields are responsible for the $U(1)^\prime$ breaking and therefore to give mass to the $Z^\prime$ gauge boson. 

The scalar potential for the fields in our framework (see Table \ref{tab:charges}) can be split in three parts,
\begin{equation}
V=V(H)+V(H,\phi_1,\phi_2)+V(\phi_1,\phi_2) \;.
\end{equation}
The first part is the SM Higgs potential, the second is the coupling of the Higgs doublet with the singlet fields,
\begin{equation}
V(H,\phi_1,\phi_2)=\lambda_2 H^\dagger H (\phi_1^*\phi_1) +\lambda_3 H^\dagger H (\phi_2^*\phi_2)  \;,
\end{equation}
and the third part is the potential for the two singlet fields,
\begin{equation}
V(\phi_1,\phi_2)=\mu_1^2\phi_1^*\phi_1+\mu_2^2\phi_2^*\phi_2+\lambda_4 (\phi_1^*\phi_1)^2+\lambda_5 (\phi_2^*\phi_2) ^2+\lambda_6 (\phi_1^*\phi_1)(\phi_2^*\phi_2) +\kappa  \phi_1\phi_1 \phi_2^*+h.c.
\end{equation}

Now once new scalar fields $\phi_i$ attain their vev ($ v_{i}/\sqrt{2} $), we get  mass for  the $Z^\prime$ gauge boson as
\begin{equation}\label{eq:BreakingU1}
\frac{1}{2}M_{Z^\prime}^2 = g^{\prime 2}~\frac{1}{2} (v^{2}_{1} + 4 v^{2}_{2} ) \;, 
\end{equation}
where we have used charges for the  $\phi_{i}  $ as mentioned in Table~\ref{tab:charges}.
Now to give an order of estimation about the breaking scale, we take mass of $  Z^{\prime} $ gauge boson 
$ M_{Z^\prime} = 0.1 $ GeV,  whereas coupling strength $ g^\prime $ is taken as $\approx 2.8 \times 10^{-5} $. 
Note that we consider these numerical values in such a way that these  can be probed in future COHERENT experiments (for a detail discussion see Sec.~\ref{sec:coherent} and Fig.~\ref{fig:mass_coupling}). 
Using these numerical values in Eq.~\eqref{eq:BreakingU1}, one finds the vevs of $\phi_1$ and $\phi_2$ as $ v_1 \approx 3 $ TeV and $ v_2  \approx 1$ TeV, respectively. It is worth to mention that the Higgs vacuum stability can be obtained when coupled to  singlet scalar field, whose vev is at the  TeV scale~\cite{Bonilla:2015kna}.

\begin{table}[t]
    \centering
    \begin{tabular}{|c|c|c|c|c|c|c|c|c|c|c|c|c|}\hline
              &$L_e$ &$L_\mu$ &$L_\tau$ & $l_e$ & $l_\mu$   &$l_\tau$  & $N_1$ & $N_2$ & $N_3$ & $H$&$\phi_1$ &$\phi_2$  \\ \hline
     $SU(2)_L$& 2 & 2 & 2 & 1 & 1 & 1 & 1 & 1 & 1 & 2 & 1 & 1 \\ \hline
      $U(1)^{\prime}$ & $x_e$ & $x_\mu$ & $x_\tau$ & $x_e$ & $x_\mu$ & $x_\tau$ & $x_e$ & $x_\mu$ & $x_\tau$ & 0 &1 & 2  \\ \hline
    \end{tabular}
    \caption{\footnotesize $U(1)^{\prime}$ charges of the model. The charges $x_\alpha$, with $\alpha=e,\mu,\tau$, can take the values $x_\alpha = 0,-1,-2$, while the charges of the quarks are $1/3$.}
    \label{tab:charges}
\end{table}

%
\begin{table}[ht]
    \centering \scriptsize
    \begin{tabular}{|c|c|c|c|c|c|} \hline
        \toprule
        $x_e$ & $x_\mu$ & $x_\tau$ & Neutrino mass matrix & Type & NSI parameters \\   \hline  \hline
        0 & -1 & -2 & $\left(\begin{array}{ccc}
        0 & 0 & \times \\
        0 & \times & \times \\
        \times & \times & \times \\
        \end{array}\right)$  & $A_1$  & $\eps_{\mu\mu}$ \& $\eps_{\tau\tau}$ \\  \hline
        0 & -2 & -1 & $\left(\begin{array}{ccc} 
        0 & \times & 0 \\
        \times & \times & \times \\
        0 & \times & \times \\
        \end{array}\right)$ & $A_2$ & $\eps_{\mu\mu}$ \& $\eps_{\tau\tau}$ \\  \hline
        -1 & 0 & -2 & $\left(\begin{array}{ccc} 
        \times & 0 & \times \\
        0 & 0 & \times \\
        \times & \times & \times \\
        \end{array}\right)$ & $B_3$ & $\eps_{ee}$ \& $\eps_{\tau\tau}$\\   \hline
        -1 & -2 & 0 & $\left(\begin{array}{ccc} 
        \times & \times & 0 \\
        \times & \times & \times \\
        0 & \times & 0 \\
        \end{array}\right)$ & $B_4$ &$\eps_{ee}$ \& $\eps_{\mu\mu}$ \\  \hline
        -2 & -1 & 0 & $\left(\begin{array}{ccc} 
        \times & \times & \times \\
        \times & \times & 0 \\
        \times & 0 & 0 \\
        \end{array}\right)$ & $\times$ &  $\eps_{ee}$ \& $\eps_{\mu\mu}$\\   \hline
        -2 & 0 & -1 & $\left(\begin{array}{ccc} 
        \times & \times & \times \\
        \times & 0 & 0 \\
        \times & 0 & \times \\
        \end{array}\right)$ & $\times$ & $\eps_{ee}$ \& $\eps_{\tau\tau}$ \\  \hline
    \end{tabular} 
    \caption{\footnotesize Neutrino mass matrix textures depending on the choices of the charges $x_\alpha$. Notice that only four of these two zero textures are allowed by the latest neutrino oscillation data~\cite{Capozzi:2016rtj,Esteban:2016qun,deSalas:2018bym}.} \label{tab:textures}    
\end{table}

The Yukawa Lagrangian  that is invariant under $SM\otimes U(1)^{\prime}$ for charged-leptons and neutrinos can be written as
\begin{eqnarray}
-\mathcal{L}_{Y} & \supset
& y_{e}\overline{L}_{e}\ell_{e}H + y_{\mu}\overline{L}_{\mu}\ell_{\mu} H + y_{\tau}\overline{L}_{\tau} \ell_{\tau} H + y_{1}^{\nu}\overline{L}_{e}\tilde{H}N_1+y_{2}^{\nu}\overline{L}_{\mu}\tilde{H}N_2+y_{3}^{\nu}\overline{L}_{\tau}\tilde{H}N_3 \;,
\label{Lag:lept1}
\end{eqnarray}
where,  $\tilde{H}=i\tau_2 H^\dagger$. 
It is clear from Eq. (\ref{Lag:lept1}) that the charged lepton mass matrix as well as the Dirac neutrino mass matrix are diagonal.

There are several anomaly-free solutions to the $U(1)$ involving baryon numbers, for scenarios where CE$\nu$NS and NSI have been explored, see for instance~~\cite{Denton:2018xmq,Heeck:2018nzc,Han:2019zkz,Papoulias:2019xaw}. In our approach, we choose the anomaly-free solution for the $U(1)^\prime=U(1)_{B-2L_\alpha-L_\beta}$, see appendix~\ref{sec:anomalycanc} for details.
In this case, let's take one of the solutions, namely $(x_e,x_\mu,x_\tau)=(0,-1,-2)$ for instance, the right-handed (RH) neutrino Lagrangian is given by
\begin{eqnarray}
-\mathcal{L}_{Majorana} & = 
& \frac{1}{2}M_1\overline{N_1^c} N_1 + \frac{1}{2} y^{N}_1 \overline{N_1^c}N_2\phi_1 + \frac{1}{2} y^{N}_2 \overline{N_1^c}N_3\phi_2+ \frac{1}{2} y^{N}_3 \overline{N_2^c}N_2\phi_2\;.
\label{Lag:lept}
\end{eqnarray}
The six possible assignments under the $U(1)^\prime$ charges for the leptons are given in Table~\ref{tab:textures} and each of the charge assignments give rise to a different model, namely different neutrino masses and mixings as well as different NSI.

Having discussed our theoretical set-up, in the subsequent sections we aim to discuss phenomenological importance of the model.
In what follows, we first examine the potential of CE$\nu$NS processes to explain NSIs as given in Table~\ref{tab:textures}. Later, predictions for neutrino oscillation parameters as well as the effective Majorana neutrino mass have been analyzed for  the allowed two-zero textures  as mentioned in Table~\ref{tab:textures}.

\section{Coherent elastic neutrino-nucleus scattering}\label{sec:coherent}

Coherent elastic neutrino-nucleus scattering has already been measured by the COHERENT experiment~\cite{Akimov:2017ade}, using a scintillator detector of made of CsI. The low energy neutrino beam was generated from the Spallation Neutron Source (SNS) at Oak Ridge National Laboratory.

The SM differential cross section for CE$\nu$NS process is given by~\cite{Drukier:1983gj,Barranco:2005yy,Patton:2012jr}
\begin{equation}
\frac{d\sigma}{dT} = \frac{G_F^2}{2\pi}M_N Q_w^2 \left(2 - \frac{M_N T}{E_\nu^2}\right),
\label{eq:crossSec}
\end{equation}
where $T$ is the nuclear recoil energy, $E_\nu$ is the incoming neutrino energy, and $M_N$ is the nuclear mass. Also, $Q_w^2$ is the weak nuclear charge and is given by
\begin{equation}
Q_w^2 = \left[ Z g_p^V F_Z(Q^2) + N g_n^V F_N(Q^2)\right]^2 \;,
\label{eq:weakCharge}
\end{equation}
where $Z(N)$ is the proton (neutron) number, $Q$ is the momentum transfer, $F_{Z(N)}(Q^2)$ its nuclear form factor,  and $g_p^V = 1/2 -2\sin^2\theta_W$, $g_n^V = -1/2$ are the SM weak couplings. It is important to notice that the cross section depends highly on the mass of the detector and the type of material, especially on the number of neutrons $N$, since the dependence on $Z$ is almost negligible due to the smallness of $g_p^V$ ($\sim0.02$). 
For the SNS energy regime, another important feature of the CE$\nu$NS cross section are the nuclear form factors.
From now onwards, we will adopt the Helm form factor~\cite{Helm:1956zz}, where equal values of proton and neutron rms radius have been used. In Table~\ref{tab:rmsValues} we present the corresponding values for different isotopes. For the analysis of the CsI detector, we will use the best-fit value of $R_n = 5.5$ from Ref.~\cite{Cadeddu:2017etk}.

\begin{table}[h]
	\centering  \scriptsize
		\begin{tabular}{|c|c|c|ccc|ccccc|} \hline
			\multicolumn{3}{|c}{} & \multicolumn{3}{|c}{Argon} & \multicolumn{5}{|c|}{Germanium} \\ \hline
			$^{23}$Na & $^{127}$I & $^{133}$Cs & $^{36}$Ar (0.33) & $^{38}$Ar (0.06) & $^{40}$Ar (99.6) & $^{70}$Ge (20.4) & $^{72}$Ge (27.3) & $^{73}$Ge (7.76) & $^{74}$Ge (36.7) & $^{76}$Ge (7.83) \\
			2.993 & 4.750 & 4.804 & 3.390 & 3.402 & 3.427 &  4.041 &  4.057 &  4.063 &  4.074 &  4.090  \\ \hline    
		\end{tabular}
		\caption{\footnotesize Proton rms radius (in fm) of the stable isotopes of sodium, iodine, cesium, argon and germanium~\cite{Angeli:2013epw}. Their percentage relative abundance is provided in parenthesis. For detectors made of argon and germanium, we  use the average $\langle R_p\rangle = \sum_i X_i R_p^i$, where $X_i$ and $R_p^i$ stand for the relative abundance and proton rms radius of the $i-$th isotope, respectively.}
		\label{tab:rmsValues}
	\end{table}

The differential recoil spectrum can be computed as
\begin{equation}
\frac{dR}{dT} = \sum_\alpha \frac{N_A M_\mathrm{det}}{M_m} \int_{E_\nu^\mathrm{min}}^{E_\nu^\mathrm{max}} \phi_\alpha (E_\nu) \frac{d\sigma}{dT} dE_\nu \;.
\label{eq:recoilSpec}
\end{equation}
Here, $N_A$ is the Avogadro's number, $M_\mathrm{det}$ is the detector mass, $M_m$ is the molar mass of the material, and $\phi_\alpha (E_\nu)$ is the neutrino flux for each flavor. The SNS neutrino flux consists of monochromatic $\nu_\mu$ coming from $\pi^+$ decays, along with delayed $\nu_e$ and $\bar{\nu}_\mu$ from the  subsequent $\mu^+$ decays. Each of these flux components are given by 
\begin{align}
	\phi_{\nu_\mu}(E_\nu) &= \eta \, \delta\left( E_\nu - \frac{m_\pi^2 - m_\mu^2}{2 m_\pi^2}\right), \\
	\phi_{\nu_e}(E_\nu) &= \eta \frac{192 E_\nu^2}{m_\mu^3} \left(\frac{1}{2} - \frac{E_\nu}{m_\mu} \right), \\
	\phi_{\bar{\nu}_\mu}(E_\nu) &= \eta \frac{64 E_\nu^2}{m_\mu^3} \left(\frac{3}{4} - \frac{E_\nu}{m_\mu} \right),
\end{align}
for neutrino energy $E_\nu \leq m_\mu / 2 \simeq 52.8$ MeV. The normalization constant is $\eta = r N_\mathrm{POT}/4\pi L^2$, where $r = 0.08$ is the fraction of neutrinos produced for each proton on target, $N_\mathrm{POT}$ represents the total number of protons on target ($\sim 2.1 \times 10^{23}$ POT) per year, and $L$ is the distance from the detector.

\begin{figure}
	\centering \includegraphics[scale=0.5]{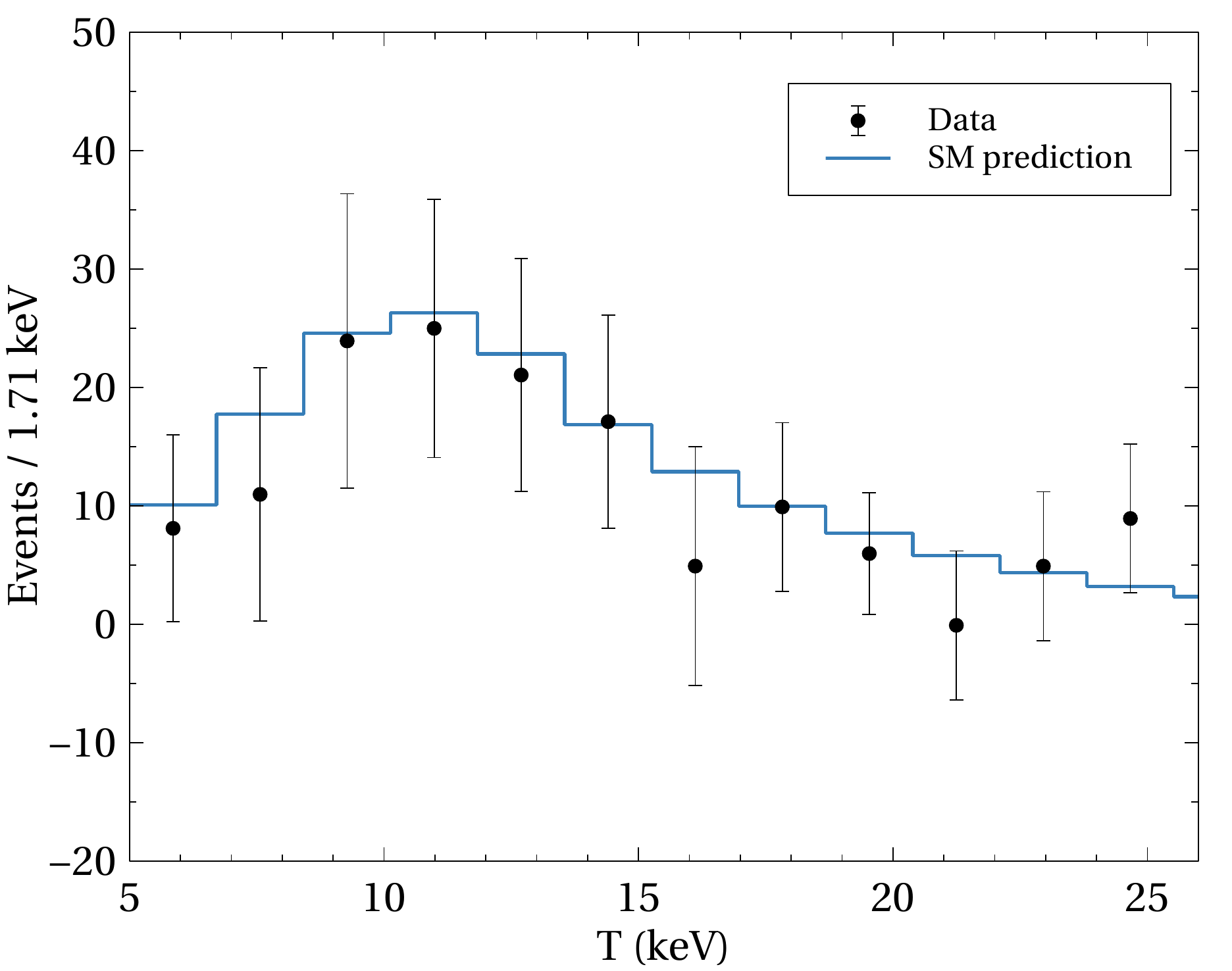}
	\caption{\footnotesize Expected number of events for the CsI detector of the COHERENT collaboration against the nuclear recoil energy $T$. The solid blue line shows the expected events in the SM framework, while the black points correspond to the experimental measurements~\cite{Akimov:2017ade}.}
	\label{fig:eventsCOHERENT}
\end{figure}

From Eq.~\eqref{eq:recoilSpec} we can compute the expected number of neutrinos per energy bin:
\begin{equation}
N_i = \int_{T_i}^{T_{i+1}}A(T)\frac{dR}{dT} dT \;,
\label{eq:expectedEvents}
\end{equation}
where $A(T)$ is the acceptance function, taken from the COHERENT data released in~\cite{Akimov:2018vzs}. In Fig.~\ref{fig:eventsCOHERENT} we show the measured number of events from the COHERENT collaboration as a function of the nuclear recoil energy T, for the expected number of events in the SM framework.


In presence of NSI, the cross section for CE$\nu$NS is affected through the weak nuclear charge (see Eq.~\eqref{eq:weakCharge}) in the following way:
\begin{equation}
Q_{w \,\alpha}^2 = \left[ Z (g_p^V + 2\eps_{\alpha\alpha}^{uV} + \eps_{\alpha\alpha}^{dV})F_Z(Q^2) + N(g_n^V  + \eps_{\alpha\alpha}^{uV} + 2\eps_{\alpha\alpha}^{dV})F_N(Q^2)\right]^2,
\label{eq:weakChargeNSI}
\end{equation}
where $\alpha = (e,\mu,\tau)$. Notice that with this new contribution, the differential cross section from Eq.~\eqref{eq:crossSec} is now flavor dependent. 

It is possible to write an effective low-energy Lagrangian for the neutrino-fermion interactions with the $Z'$ boson as
\begin{equation}\label{eq:eff_lag}
\mathcal{L}_\mathrm{eff} = -\frac{g'^2}{Q^2 + M_{Z'}^2}\left[ \sum_\alpha x_\alpha \bar{\nu}_\alpha \gamma^\mu P_L \nu_\alpha \right] \left[ \sum_q x_q \bar{q}\gamma_\mu q \right],
\end{equation}
where $Q^2$ is the transferred momentum. Therefore, by comparing this effective Lagrangian with the NSI Lagrangian in Eq.~\eqref{eq:NSI}, we can relate the NSI parameters with the $Z'$ interaction parameters as
\begin{equation}
\eps_{\alpha\alpha}^{qV} = \frac{g'^2 x_\alpha x_q}{\sqrt{2}G_F (Q^2 + M_{Z'}^2)}\;.
\label{eq:nsiCouplingMass}
\end{equation}


\begin{figure}[h]
\centering
\includegraphics[height=4.5cm, width=16cm]{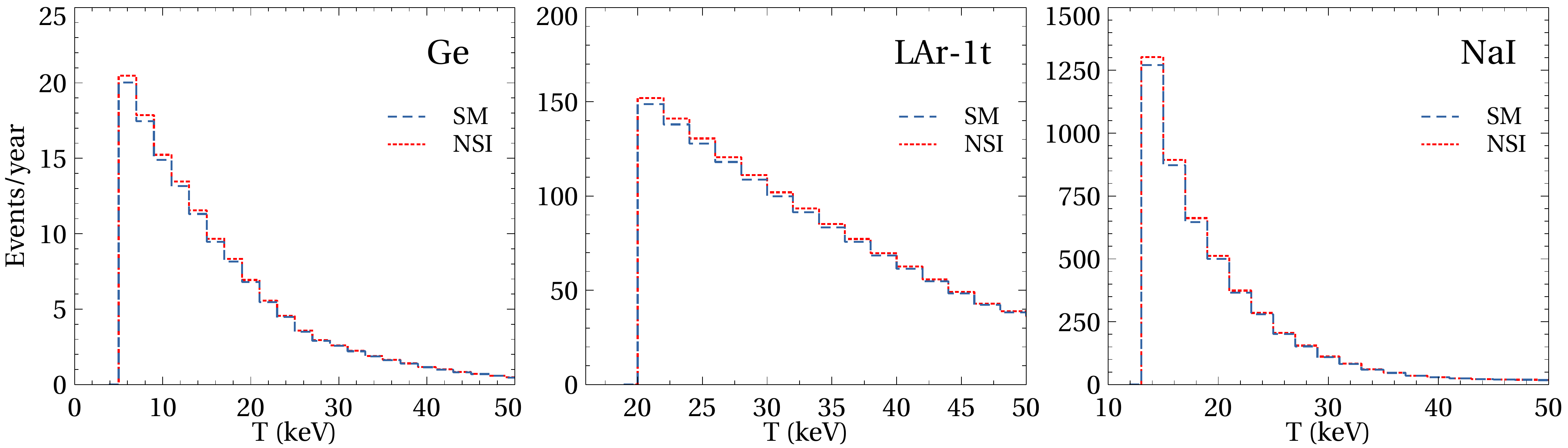}
	\caption{\footnotesize Expected number of events per year as a function of the nuclear recoil energy for the upcoming Ge, LAr-1t, and NaI detectors as shown in the left, middle, and right panel, respectively. The blue dashed lines correspond to the SM framework, while the red dotted lines represent the NSI scenario with $M_{Z'} = 0.1$ GeV and $g' = 2\times 10^{-5}$, for the type $A_2$ model. The details about the benchmark values of $M_{Z^\prime}$ and $ g^\prime $ that are considered here are presented in Fig.~\ref{fig:mass_coupling}.}
\label{fig:eventsFutureCOHERENT}
\end{figure}

\begin{table}[h]
	\centering  \scriptsize
	\begin{tabular}{|c|c|c|c|c|} \hline
		Detector & Mass (kg) & Baseline (m)  & Energy threshold (keV) & Efficiency \\ \hline \hline
		CsI~\cite{Akimov:2017ade}  & 14.6 & 19.3& 5 & $A(T)$~\cite{Akimov:2018vzs} \\  \hline
		CENNS-10~\cite{Akimov:2020pdx}  & 24 & 27.5& 20 & $F(T)$~\cite{Akimov:2020pdx} \\  \hline
		LAr-1t~\cite{Akimov:2018ghi}  & 610  & 29  & 20 & 0.5 \\ \hline
		Ge~\cite{Akimov:2018ghi}   & 10   & 22  & 5 & 0.5 \\ \hline
		NaI~\cite{Akimov:2018ghi}  & 2000 & 28  & 13 & 0.5\\ \hline
	\end{tabular}
	\caption{\footnotesize Specifications of the current COHERENT-CsI~\cite{Akimov:2017ade} and CENNS-10~\cite{Akimov:2020pdx} detectors, along with the future setups using other type of detectors~\cite{Akimov:2018ghi}. For CENNS-10, the efficiency function $F(T)$ is taken from Fig. 3 of Ref.~\cite{Akimov:2020pdx} for the analysis B. Since there is no information about the efficiencies of the future detectors (viz, LAr-1t, Ge, and NaI), we have assumed a conservative flat efficiency of $50\%$.}
	\label{tab:prospects}
\end{table}

\begin{figure}[t]
	\centering 
	\includegraphics[scale=0.5]{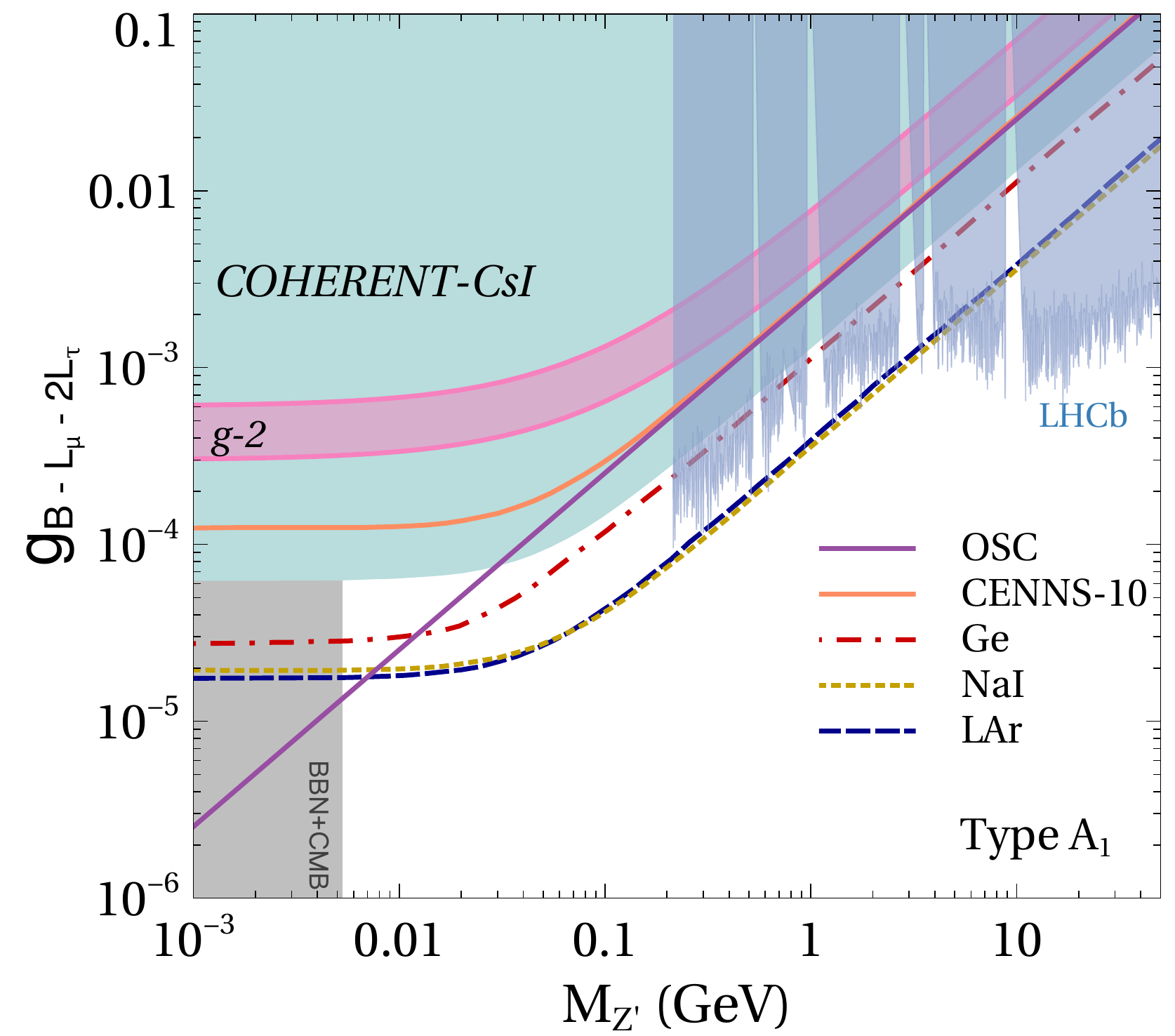}%
	\includegraphics[scale=0.5]{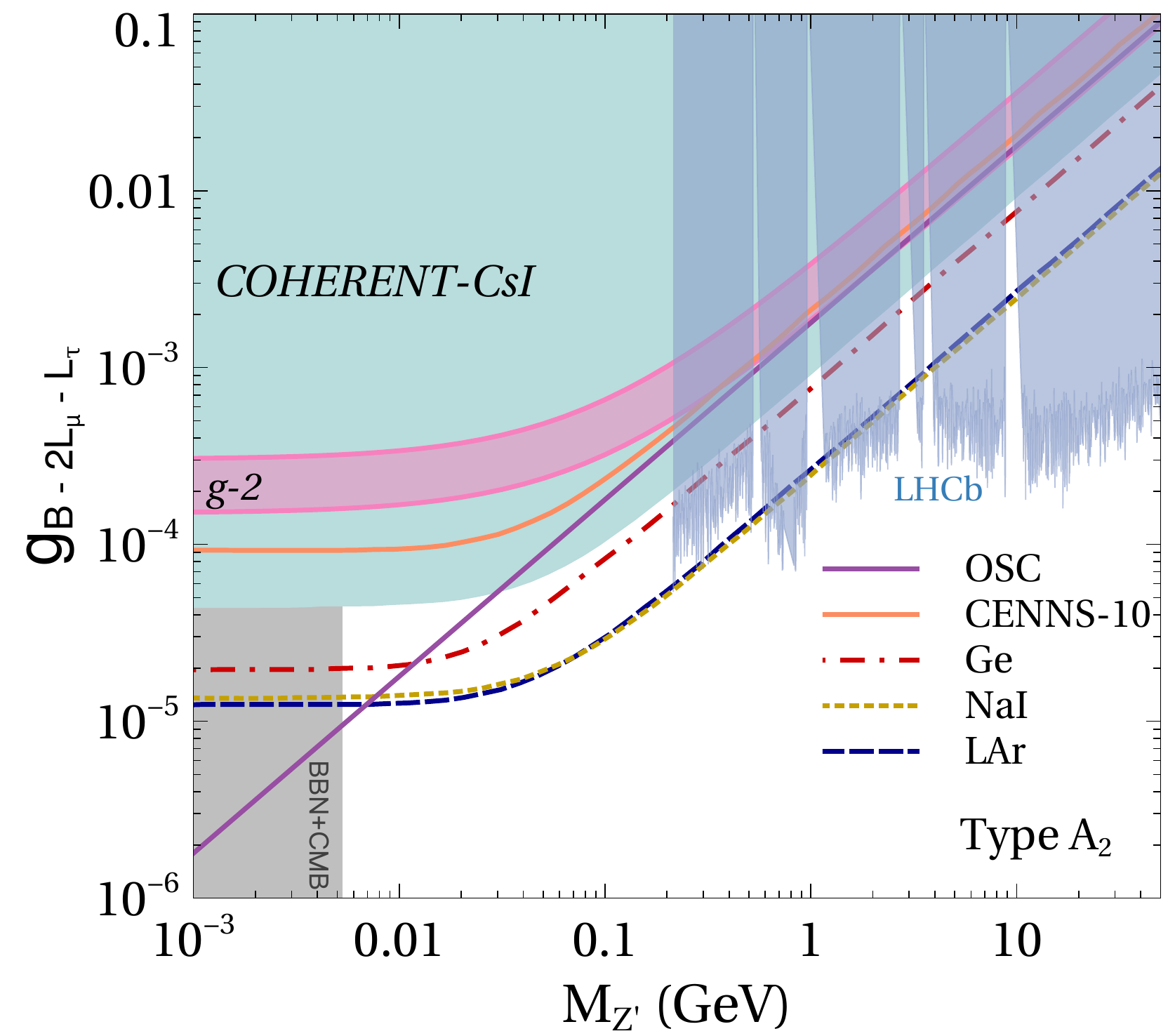}
	\includegraphics[scale=0.5]{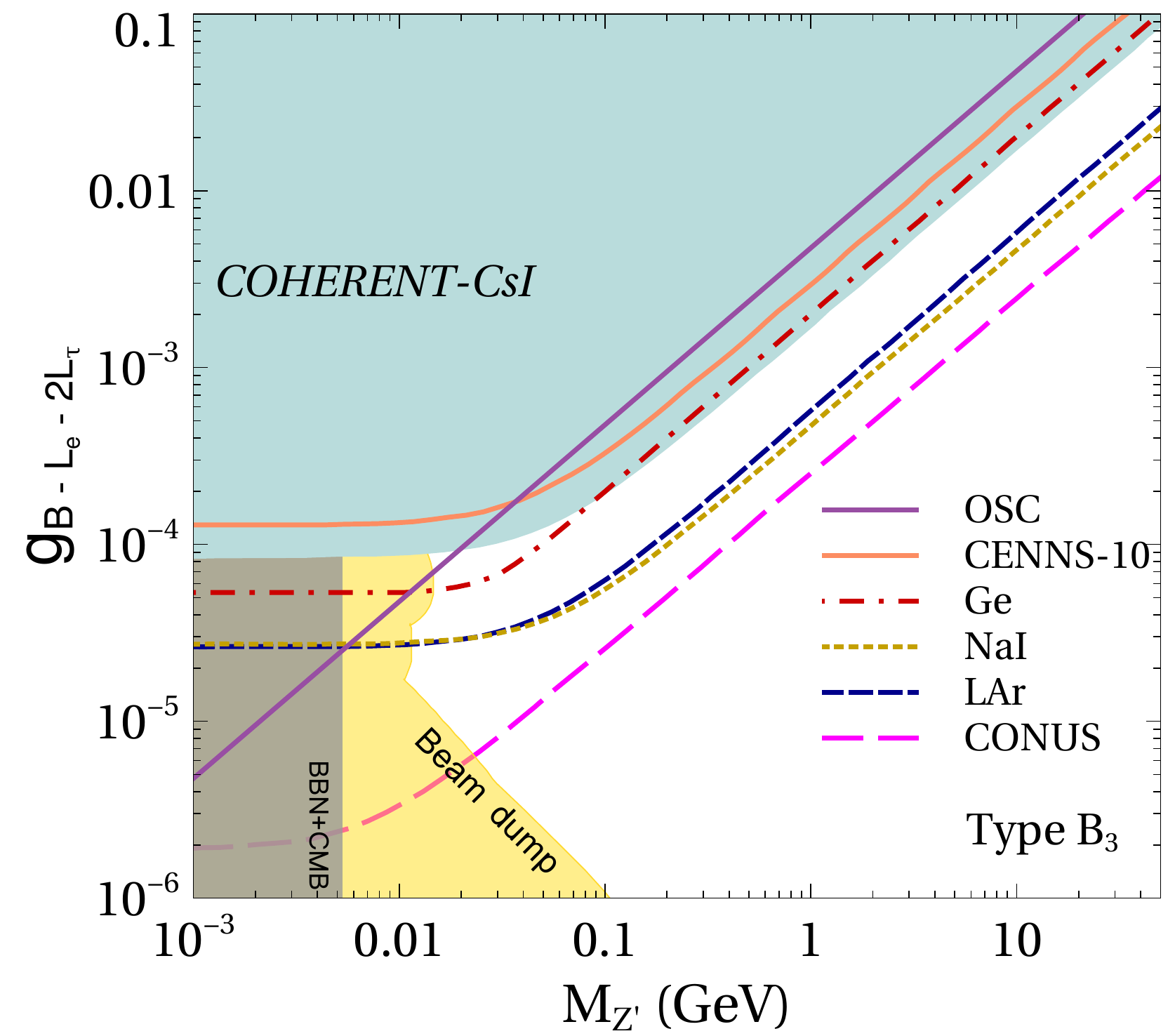}%
	\includegraphics[scale=0.5]{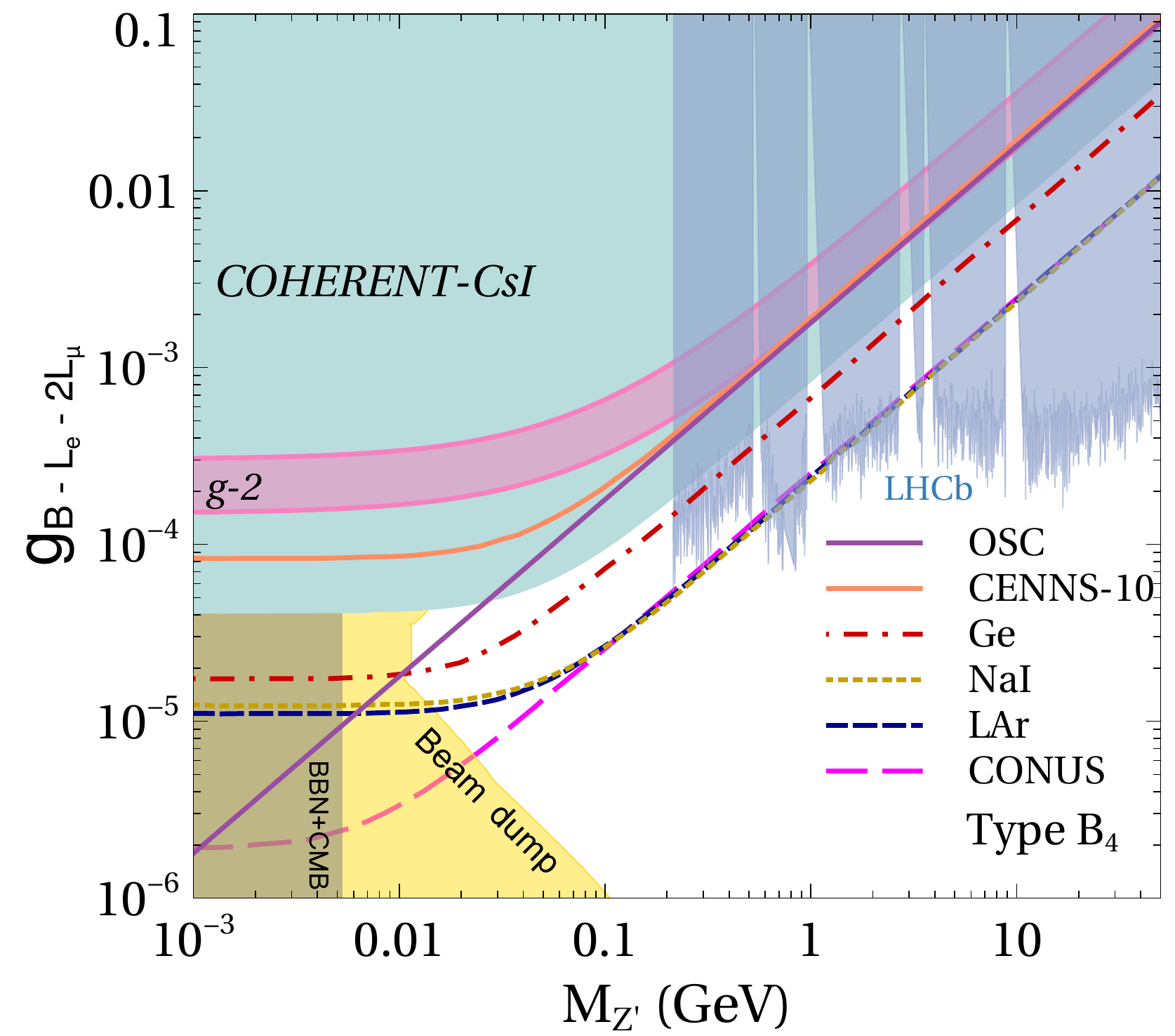}
	\caption{\footnotesize Exclusion regions at $95\%$ C.L. in the $(M_{Z'}, g')$ plane for the different models. The light-green shaded area corresponds to the constraint set by the current COHERENT measurement using a CsI detector~\cite{Akimov:2017ade}, while the orange solid line comes from the recent COHERENT results using the CENNS-10 detector~\cite{Akimov:2020pdx}. The solid purple line shows the limit from oscillation experiments~\cite{Esteban:2018ppq}. The limits set by the future detectors setup from the COHERENT collaboration~\cite{Akimov:2018ghi},  namely, Ge, NaI, and LAr-1t are shown using the red dash-dotted, yellow dotted, and blue dashed lines, respectively.
The limits from the  CONUS reactor experiment~\cite{Lindner:2016wff} are shown by the  magenta (long dashed) lines. The exclusion regions set by the beam dump experiments~\cite{Bergsma:1985qz,Tsai:2019mtm,Bjorken:1988as,Riordan:1987aw,Bross:1989mp,Konaka:1986cb,Banerjee:2019hmi,Astier:2001ck,Davier:1989wz,Bernardi:1985ny}, BBN and CMB~\cite{Kamada:2015era}, and LHCb dark photon searches~\cite{Aaij:2019bvg} are presented using color code yellow, gray and sky-blue regions, respectively. The pink shaded band corresponds to the region where the muon $(g-2)$ anomaly is explained~\cite{Jegerlehner:2009ry} (see text for more details).}
	\label{fig:mass_coupling} \label{fig:coherent}
\end{figure}

In Fig.~\ref{fig:eventsFutureCOHERENT}, we plotted the number of events versus the nuclear recoil energy, for different detectors (Ge, LAr-1t and NaI) considering  the future plans of the COHERENT collaboration. The features of the future detectors that are used in our numerical simulations, along with the current CsI detector are presented in Table~\ref{tab:prospects}. We show the expected events in the SM framework, and compare with the case of NSI terms in the cross section. For this particular example, we considered the  $ A_2 $ model with $M_{Z'} = 0.1$ GeV and $g' = 2\times 10^{-5}$.
We give  the details about the  values of $M_{Z^\prime}$ and $ g^\prime $ that are considered here in 
Fig. \ref{fig:mass_coupling}. As expected, the number of events in presence of NSI increases with respect of those in the SM, but this increase is higher for smaller values of the nuclear recoil energy $T$.

Given the relation in Eq.~\eqref{eq:nsiCouplingMass}, it is now clear how the NSIs can be generated from the interactions of a new vector boson $Z'$. By computing the number of events including NSI contributions, we are now able to compare with the COHERENT measurements in order to set boundaries to the coupling and mass of the $Z'$ boson. 

As mentioned before, the first part of the analysis consists in comparing with the first measurements of CE$\nu$NS, provided by the COHERENT collaboration~\cite{Akimov:2017ade}. A CsI detector of 14.6 kg was used at a distance of 19.3 m from the source. The cross section for this type of detector has to be computed separately for cesium (Cs) and iodine (I) in the following way:
\begin{equation}
\frac{d\sigma}{dT}= \left(\frac{d\sigma}{dT}\right)_\mathrm{Cs} + \left(\frac{d\sigma}{dT}\right)_\mathrm{I}.
\end{equation}

We perform a fit of the COHERENT-CsI data by means of a least-squares function
\begin{equation}
\chi^2 = \sum_{i=4}^{15} \left[\frac{N_\mathrm{meas}^i - (1+\alpha)N_\mathrm{th}^i- (1+\beta)B_\mathrm{on}^i}{\sigma_\mathrm{stat}^i} \right]^2 + \left(\frac{\alpha}{\sigma_\alpha}\right)^2 + \left(\frac{\beta}{\sigma_\beta}\right)^2,
\label{eq:chiSquareFunction}
\end{equation}
where $N_\mathrm{meas}^i (N_\mathrm{th}^i)$ is the measured (expected) number of events per energy bin, $\sigma_\mathrm{stat}^i = \sqrt{N_\mathrm{th}^i + B_\mathrm{on}^i + 2B_\mathrm{ss}^i}$ is the statistical uncertainty. Also, $B_\mathrm{on}^i$ and $B_\mathrm{ss}^i$ are the beam-on and steady-state backgrounds, respectively. We marginalize over the nuisance  parameters $\alpha$ and $\beta$, which quantify to the signal and background normalization uncertainties $\sigma_\alpha$ and $\sigma_\beta$, respectively. Following the COHERENT-CsI analysis, we choose $\sigma_\alpha = 0.28$, which includes neutrino flux ($10\%$), signal acceptance ($5\%$), nuclear form factor ($5\%$) and quenching factor ($25\%$) uncertainties, and $\sigma_\beta=0.25$~\cite{Akimov:2017ade}. Since the fit to the quenching factor was done for the bins from $i=4$ to $15$, we follow our analysis  only for these energy bins.

In order to extract  information about the $Z'$ boson, we compute the expected number of events $N_\mathrm{th}$ including NSI effects, according to Eq.~\eqref{eq:expectedEvents} and the weak nuclear charge in Eq.~\eqref{eq:weakChargeNSI}. It must be pointed out that in the NSI scenario, the differential cross section is now flavor dependent.

As we have mention before (see section \ref{sec:model} for  details), the proposed model has six possibilities depending on the $U(1)'$ charges of the charged leptons. Since only four of these cases are allowed by oscillations data ($A_1$, $A_2$, $B_3$ and $B_4$), we will perform the $\chi^2$ analysis only for these cases. 
Note that we will give a detailed phenomenological consequences of these four two-zero textures within the standard three-flavor neutrino oscillation paradigm in the next section. 

Since all the quarks have  same $U(1)'$ charge, we get $\eps_{\alpha\alpha}^{uV} = \eps_{\alpha\alpha}^{dV}$, reducing the number of free parameters. Also, the neutrino source does not produce tau neutrinos, and hence, we can not extract any information about $\eps_{\tau\tau}$.

It is to be noted that the COHERENT collaboration has reported the first measurement of \cevns with argon by using the CENNS-10 detector, which corresponds to $13.7\times 10^{22}$ POT. The CENNS-10 detector has an energy threshold of 20 keV, an active mass of 24 kg, and is located at 27.5 m from the SNS target. As described in Ref~\cite{Akimov:2020pdx}, the collaboration has performed two independent analyses, labeled A and B. Analysis B yielded a total of 121 \cevns events, 222 beam-related and 1112 steady-state background events.

To extract exclusion regions for the $Z'$ parameters, we perform a single-bin analysis, using a $\chi^2$ function equivalent to Eq.~\eqref{eq:chiSquareFunction}. Following the analysis B of Ref~\cite{Akimov:2020pdx}, we take $\sigma_\alpha = 0.07$ and $\sigma_\beta = 0.107$. For the calculation of the number of events, we use the efficiency function provided in Fig. 3 from Ref~\cite{Akimov:2020pdx}.

In Fig.~\ref{fig:mass_coupling}, we show the exclusion regions at $95\%$ C.L. in the $(M_{Z'}, g')$ plane. Each panel corresponds to one of the four possible models, where the resulting light neutrino mass matrix is of type $A_1, A_2, B_3$ and $B_4$. The constraints coming from the COHERENT data, using the CsI and CENNS-10 detectors, has been presented using the light-green shaded region and the orange solid line, respectively.
In order to have a more complete study, we also include exclusion regions arising from the future upgrades of the COHERENT collaboration: Ge,  NaI, and LAr-1t detectors, considering  a $10\%$ SM signal as background and an exposure of four years. For this analysis, we consider a decrease in the quenching factor uncertainty by a factor of two with respect to the CsI detector case ($12.5\%$). This improvement leads to a signal nuisance parameter of $\sigma_{\alpha}=0.175$, while the background  parameter remains the same $\sigma_\beta = 0.25$. We show the exclusion regions using the red dash-dotted, yellow dotted, and blue dashed lines, respectively.
We can see how these future setups can improve the current COHERENT limits for the coupling $g'$ by almost one order of magnitude. 

Notice that the propagation of neutrinos  in matter are affected by coherent forward scattering where one have zero momentum transfer. Hence, the effective Lagrangian from Eq.~(\ref{eq:eff_lag}) that is relevant for NSI can be written as
\begin{equation}\label{eq:eff_lag_new}
\mathcal{L}_\mathrm{eff} = -\frac{g'^2}{ M_{Z'}^2}\left[ \sum_\alpha x_\alpha \bar{\nu}_\alpha \gamma^\mu P_L \nu_\alpha \right] \left[ \sum_q x_q \bar{q}\gamma_\mu q \right],
\end{equation}
%
%
irrespective of the $Z^\prime$ mass. In this limit Eq.~(\ref{eq:nsiCouplingMass}) becomes
\begin{equation}
\eps_{\alpha\alpha}^{qV} = \frac{g'^2 x_\alpha x_q}{\sqrt{2}G_F (M_{Z'}^2)} \;.
\label{eq:nsiCouplingMass1}
\end{equation}
%
%
%
In Fig.~\ref{fig:mass_coupling}, we also include limits coming from 
oscillation experiments (see purple solid line) using the relation given in Eq.~(\ref{eq:nsiCouplingMass1}).
%
%
For models $A_1$ and $A_2$, we take the smallest value of $\eps_{\mu\mu}$ from the first column of Table~\ref{tab:nsiConstraints}, when setting $\eps_{ee}=\eps_{\tau\tau}=0$. Then we use Eq.~\eqref{eq:nsiCouplingMass1} to get a limit for $g'$ as a function of $M_{Z'}$. For $B_3$, we extract a value for $\eps_{ee}$ by taking $\eps_{\mu\mu}=0$ \footnote{For model $B_4$ we considered the smallest possible value between $\eps_{ee}$ and $\eps_{\mu\mu}$ from the other models.}. The limit from BBN +CMB~\cite{Kamada:2015era} is also presented using gray band in Fig.~\ref{fig:mass_coupling}. For the cases where $\eps_{ee}\neq 0$ (i.e., for $B_3$ and $B_4$), we have also included boundaries for a light $Z'$ boson, obtained by different electron beam dump experiments as shown by the yellow region. We have used the \texttt{Darkcast}~\cite{Ilten:2018crw} code to translate the beam dump limits to our specific model.
In the cases where $\eps_{\mu\mu}$ is present, we also consider limits set by dark photon searches for LHCb limits~\cite{Aaij:2019bvg} shown using the sky-blue region. 
 We also use the \texttt{Darkcast}~\cite{Ilten:2018crw} code to translate these limits to the different cases of our model, which  has been shown using the sky-blue regions in Fig.~\ref{fig:mass_coupling}.

 The interaction of the $Z'$ boson with muons leads to an additional contribution to the anomalous magnetic moment:
\begin{equation}
\delta a_\mu =\frac{g'^{\,2} x_\mu^2}{8\pi^2} F\left(\frac{M_{Z'}}{m_\mu}\right) \;,
\end{equation} 
where
\begin{equation}
F(x) = \int_0^1 dz \frac{2z(1-z)^2}{(1-z)^2 + x^2z^2} \;.
\end{equation}

Since the existence of new light vector bosons can explain the inconsistency in the anomalous magnetic moment of the muon, $(g-2)_\mu$, ~\cite{Gninenko:2001hx,Baek:2001kca}, we have incorporated boundaries arising from this process in Fig.~\ref{fig:mass_coupling}. The region of the  $(M_{Z'}, g')$ plane where our model can explain the discrepancy $\Delta a_\mu = (29 \pm 9)\times 10^{-10}$~\cite{Jegerlehner:2009ry} is the pink region.
Notice that only in the $B_3$  this region is absent, since there is no interaction between muons and the $Z'$ boson ($\eps_{\mu\mu} = 0$).

Furthermore, there are several proposals aiming to measure CE$\nu$NS using nuclear reactors, such as CONNIE~\cite{Aguilar-Arevalo:2016qen}, CONUS~\cite{Lindner:2016wff}, MINER~\cite{Agnolet:2016zir}, RED100~\cite{Akimov:2012aya}, TEXONO~\cite{Wong:2005vg}, etc. For example, the CONUS experiment will consist of a 4 kg Germanium detector with an energy threshold of $300$ eV, located at 17 m from the nuclear power plant at Brokdorf, Germany~\cite{Lindner:2016wff}. They expect $\sim10^5$ events over a 5 year run, assuming the SM signal.

We also present limits for the $Z'$ boson considering the CONUS experiment. For the calculation of the number of \cevns events, we have taken into account an antineutrino energy spectrum coming from the fission products $^{235}$U, $^{238}$U, $^{239}$Pu and $^{241}$Pu~\cite{Mueller:2011nm}. For energies below $2$ MeV, we use the theoretical results obtained in Ref.~\cite{Kopeikin:1997ve}. Since reactor antineutrinos are produced with energies of a few MeV, the nuclear form factors play no role in the detection of \cevns events, therefore we safely take them to be equal to one.
	
For this analysis, we assume a flat detector efficiency of $50\%$, and the same $\chi^2$ function given by Eq.~\eqref{eq:chiSquareFunction} with a background equal to $10\%$ of the SM signal, where uncertainties $\sigma_\alpha = 0.1$ and $\sigma_\beta = 0.25$ have been used. Since a nuclear reactor produces only electron antineutrinos, we  give an exclusion regions only for the cases where $\eps_{ee}\neq0$  ({\it i.e.} for $B_3$ and $B_4$). These regions are shown in the lower panels of Fig.~\ref{fig:mass_coupling}, denoted with the magenta dashed line.


\begin{table}[ht]
	\begin{center} \scriptsize
		\begin{tabular}{c|c|c|c} \hline
			Texture & Experiments & $ g^{\prime} (\times 10^{-5})$ & $ v_1  $ (TeV)  \\ \hline \hline
			\multirow{5}{*}{$ A_1 $}
			& Osc & 25  &  0.39 \\ 
			& COHERENT-CsI & 18 & 0.55\\
			& COHERENT-Ge & 11 & 0.91\\ 
			& COHERENT-LAr-1t & 4.3 & 2.32\\
			& COHERENT-NaI & 4.1 & 2.44\\  
			\hline \hline 
			\multirow{5}{*}{$ A_2 $} 
			& Osc & 17 & 0.58\\ 
			& COHERENT-CsI & 13 & 0.77\\
			& COHERENT-Ge & 8.3 & 1.20\\ 
			& COHERENT-LAr-1t & 2.9 & 3.45\\
			& COHERENT-NaI & 2.9 & 3.45 \\
			\hline \hline 
			\multirow{5}{*}{$ B_3 $}& Osc & 47 &  0.21\\ 
			& COHERENT-CsI & 34 & 0.29\\
			& COHERENT-Ge & 19 & 0.53\\ 
			& COHERENT-LAr-1t & 6.3 & 1.58\\
			& COHERENT-NaI & 5.6 & 1.78\\
			& CONUS & 2.6 & 3.85\\
			\hline \hline 
			\multirow{5}{*}{$ B_4 $} 
			& Osc &  18 & 0.55\\ 
			& COHERENT-CsI & 10 & 0.99\\
			& COHERENT-Ge & 7.3 & 1.39\\ 
			& COHERENT-LAr-1t & 2.6 & 3.85\\
			& COHERENT-NaI & 2.6 & 3.85\\
			& CONUS & 2.6 & 3.85\\
			\hline                           
		\end{tabular}
		\caption{\footnotesize Set of representative values for the coupling constant $  g^{\prime} $ and for the symmetry breaking scale  $ v_1  $ have been provided for different detectors and experimental limits as shown in Fig. \ref{fig:mass_coupling}. Here we fix $ M_{Z^{\prime}} = 100 $ MeV and $ v_2  = 1$ TeV. 
		}
	\end{center}
	\label{tab:gPrimeLimit}
\end{table}



The first panel of Fig.~\ref{fig:mass_coupling}, i.e., $A_1$ ($U(1)_{B-L_\mu-2L_{\tau}}$) has $\epsilon_{\mu\mu}$ and $\epsilon_{\tau\tau}$ with $\epsilon_{\tau\tau}>\epsilon_{\mu\mu}$. In this scenario, it can be seen that the future COHERENT experiment with LAr-1t detector will explore a parameter space for masses between $7$ MeV to $3$ GeV and couplings as small as $g^\prime\sim 10^{-5}$. For masses between $200$ MeV and $4$ GeV the future COHERENT bounds will be competitive with the current  LHCb exclusion limits. However, we notice that above  $3$ GeV bounds coming from the LHCb drak-photon searches will give the strongest constraints, where $ g^\prime $ can be $ \sim  10^{-3}$ (see sky-blue region).   Bounds arising from the calculation of $ \Delta N_{eff} $ of BBN will rule out $M_{Z^\prime} < 7$ MeV as shown by the gray band. 
We now proceed to discuss our results for $A_2$ ($U(1)_{B-2L_\mu-L_{\tau}}$) as shown by the second panel of the first row of Fig.~\ref{fig:mass_coupling}. It has $\epsilon_{\mu\mu}$ and $\epsilon_{\tau\tau}$ as in $A_1$ but in this case $\epsilon_{\tau\tau}<\epsilon_{\mu\mu}$. Here, we have found that the future COHERENT experiment will explore a parameter space for masses between  $7$ MeV to $0.55$ GeV  and couplings up to $g^\prime\sim 10^{-5}$.  For masses between $200$ and $500$ MeV the future COHERENT bounds will be comparable as  exclusion coming from LHCb. Unlike $A_1$, LHCb can explore more parameter space for this scenario, i.e., $M_{Z^\prime}\geq 0.55$ GeV, compared to COHERENT-LAr-1t bounds.  $ \Delta N_{eff} $ also shows similar bounds as $A_1$.

Unlike the scenarios $A_1$  and $A_2$, we also have contributions coming from the beam dump experiments and reactor experiment CONUS that is because of non-zero $\epsilon_{ee}$ for $B_3$ and $B_4$, which we show at the second row 
of Fig.~\ref{fig:mass_coupling}, respectively.
The model $B_3$ ($U(1)_{B-L_e-2L_{\tau}}$) predicts NSI parameters like $\epsilon_{ee}$ and $\epsilon_{\tau\tau}$ (see Table \ref{tab:textures} for details).
 It has been observed that CONUS shows the most stringent constraint, compared to the future COHERENT-LAr-1t bounds,  for the masses greater than $25$ MeV with the coupling constant $g^\prime\sim 5\times10^{-6}$, as shown by the magenta dashed line.
Moreover, the region  $M_{Z^\prime} < 25$ MeV and $g^\prime < 5\times10^{-6}$ is ruled out by the beam dump bounds (see light-yellow region).
In our final scenario, i.e., $B_4$ ($U(1)_{B-L_e-2L_{\mu}}$), the contribution from the LHCb is also observed because of non-zero $\epsilon_{\mu\mu}$ together with $\epsilon_{ee}$.
%
We notice that for masses greater $25$ MeV up to $\sim 500$ MeV and couplings $ g^\prime $ in the range $(5\times10^{-6} - 0.5\times10^{-4})$, CONUS will show the strongest exclusion region, whereas masses $\geq 500$ MeV will be explored by LHCb. On the other hand, predictions below $M_{Z^\prime} < 25$ MeV remains same as $B_3$.

It is worth to mention that the exclusion region coming from the recent results of the  CENNS-10 detector is weaker than the future upgrade LAr-1t detector for two main reasons: the greater mass of the latter ($\sim25$ times bigger) and the total exposure that has been considered in this work (4 years).

 Finally, by investigating all the four scenarios, it has been seen that the bounds arising from $ (g-2)_{\mu} $ (see the pink band ) is ruled out by the current COHERENT-CsI data, while  limits from oscillation experiments (as shown by the solid purple line) will be ruled out by the future COHERENT data.
Finally, we present  a set of benchmark values that can be explored by different experiments in the  Table \ref{tab:gPrimeLimit}.

So far we have discussed the importance of CE$\nu$NS processes to investigate NSIs for all the possible allowed cases for the given $U(1)^{\prime}$ charges as given by Table~\ref{tab:textures}. Our next section is devoted to the predictions for the  standard three flavor neutrino oscillation parameters  as well as for the effective Majorana neutrino mass within the formalism of two-zero textures that are appeared in this gauge extended model (see Table~\ref{tab:textures}  for allowed possibilities). 

\section{Two-zero textures}\label{sec:two-zero}
%
Here we revisit the phenomenology of the two-zero textures that are allowed in this model, as given in Table~\ref{tab:textures}, viz  $A_1$, $A_2$, $B_3$, and $B_4$ in light of the latest global-fit data.
The two-zero textures  that were classified in \cite{Frampton:2002yf} are phenomenologically very appealing in the sense that they guarantee the calculability of the neutrino  mass matrix $M_{\nu}$ from which both the neutrino mass spectrum and the flavor mixing pattern can  be determined
\cite{Xing:2002ta,Desai:2002sz,Meloni:2014yea,Alcaide:2018vni}.
In what follows, we first parameterize $M_{\nu}$ in terms of the
three neutrino mass eigenvalues ($m_{1}$, $m_{2}$, $m_{3}$) and the three
neutrino mixing angles ($\theta_{12}$, $\theta_{23}$,
$\theta_{13}$) together with the three CP violating phases ($\delta$, $\alpha$, $\beta$). 
Note that here $\delta$ is the Dirac type CP-phase, whereas $\alpha$, and $\beta$ are the Majorana type CP-phases.
Therefore, the  mass matrix $M_{\nu}$ can be diagonalized by a complex unitary matrix $U$ as
\begin{equation} \label{eq:mnu}
M_{\nu} = U m_{\nu}^{diag} U^{T} \;,
\end{equation}
where $m_{\nu}^{diag} = diag \{m_1,m_2,m_3\}$. In the standard PDG formalism, the neutrino mixing
matrix $U$, also known as the PMNS matrix is given by
\begin{align} \label{eq:pmns}
U & \equiv V P  \;, \nonumber \\
& = \left(
\begin{array}{ccc}
c_{12}c_{13} & s_{12}c_{13} & s_{13}e^{-i\delta} \\
-s_{12}c_{23}-c_{12}s_{23}s_{13}e^{i\delta} &
c_{12}c_{23}-s_{12}s_{23}s_{13}e^{i\delta} & s_{23}c_{13} \\
s_{12}s_{23}-c_{12}c_{23}s_{13}e^{i\delta} &
-c_{12}s_{23}-s_{12}c_{23}s_{13}e^{i\delta} & c_{23}c_{13}
\end{array}
\right)\left(
\begin{array}{ccc}
1 & 0 & 0 \\ 0 & e^{i\alpha} & 0 \\ 0 & 0 & e^{i(\beta+\delta)}
\end{array}
\right) \;,
\end{align}
where $s_{ij} = \sin\theta_{ij}$ and $c_{ij}=\cos\theta_{ij}$. 
Given the parameterization of $U$, it is now straight forward to write down the elements of neutrino mass matrix $M_{\nu}$ with the  help of Eq.~\eqref{eq:mnu}.
%

The two-zero textures of  the neutrino mass matrix $M_{\nu}$ (see Eq.~\eqref{eq:mnu}) satisfies two complex equations as 
\begin{equation}
m_{ab} = 0,~~  m_{pq}=0 \;,
\end{equation}
where $a$, $b$, $p$ and $q$ can take values $e$, $\mu $ and
$\tau$. Above equations can also be written as
\begin{align}\label{eq:tex-sol}
m_{1}V_{a1}V_{b1} + m_{2}V_{a2}V_{b2}e^{2i\alpha
} + m_{3}V_{a3}V_{b3}e^{2i(\beta +\delta )} & = 0 \;,  \nonumber \\ 
m_{1} V_{p1}V_{q1} + m_{2}V_{p2}V_{q2}e^{2i\alpha
} + m_{3}V_{p3}V_{q3}e^{2i(\beta +\delta )} & = 0\;,
\end{align}
where $V$ has been defined in Eq.~\eqref{eq:pmns}. 
We notice that these two equations involve nine physical parameters $m_{1}$, $m_{2}$ , $m_{3}$,
$\theta _{12}$, $\theta _{23}$, $\theta _{13}$ and CP-violating
phases $\alpha $, $\beta $, and $\delta $.
The three mixing angles
$(\theta _{13}, \theta _{12}$, $\theta _{23})$ and two mass-squared differences
$(\Delta m_{12}^{2}$, $\Delta m_{23}^{2})$ are known from the
neutrino oscillation data. 
Note here that from the latest global-fit results, we have some predictions about the CP-violating
phase $\delta $, however at $ 3\sigma $,  full range i.e.,  $0^\circ - 360^\circ$ is still allowed. Therefore, in this study we kept $\delta $ as a free parameter.
The masses $m_{2}$ and $m_{3}$ can be calculated from the known mass-squared
differences $\Delta m_{12}^{2}$ and $\Delta m_{23}^{2}$ using the
relations
$m_{2}=\sqrt{m_{1}^{2}+\Delta m_{12}^{2}} \;, {\rm and}  ~~~
m_{3}=\sqrt{m_{2}^{2}+\Delta m_{23}^{2}}.$
Thus, we have two complex equations relating four unknown parameters viz. $m_1$, $\alpha$, $\beta$ and $\delta$. Therefore, one can have the predictability of all these four parameters within the formalism of two-zero textures.


We numerically solve Eq.~\eqref{eq:tex-sol} for the concerned types of two-zero textures, see Table~\ref{tab:textures}. It has been known  from the latest global analysis of neutrino oscillation results~\cite{Capozzi:2016rtj,Esteban:2016qun,deSalas:2018bym} that the least unknown parameter among  the  three mixing angles is the atmospheric mixing angle  $\theta _{23} $. 
Therefore, considering some benchmark values of $\theta _{23} $, we calculate remaining unknown parameters, which we present in Table~\ref{tab:NuParamNH} and \ref{tab:NuParamQD}. We take the latest best-fit value of $\theta _{23} $ from \cite{deSalas:2018bym} as one of our benchmark value, whereas maximal value of $\theta _{23} $ i.e., $\theta _{23} = 45^\circ$ is taken as the second benchmark value.
Notice that the seed point $\theta_{23} = 45^\circ$ has the great importance in perspective of flavor symmetries as well as flavor models building. Among  numerous theoretical frameworks, $ \mu-\tau $ symmetry that explains $\theta_{23} = 45^\circ$ has received great attention in the neutrino community, for the latest review see Ref.~\cite{Xing:2015fdg}.
From Table~\ref{tab:NuParamNH}, we notice that the textures $ A_1, A_2 $ can explain both the latest best-fit as well as the maximal value of $ \theta_{23} $. Further,  given these benchmark values we 
calculate unknown parameters $m_1$, $\alpha$, $\beta$ and $\delta$. It is to be noted from the fourth column that  the predicted values of $ \delta $ for all the cases lies within $ 1\sigma $ of the  latest best-fit value~\cite{deSalas:2018bym}, which is $237.6^{+37.8^\circ}_{-27.0^\circ}$. 
We also calculate $  m_2, m_3 $ (see second column of the Table~\ref{tab:NuParamNH}) to find $ \sum m_{\nu} $. 
From the third column, one can find that the measured values of $ \sum m_{\nu} $ for all the cases are well within the latest value provided by \textit{Planck} collaboration~\cite{Aghanim:2018eyx} which gives $ \sum m_{\nu} < 0.12$ eV (95\%, \textit{Planck} TT, TE, EE + lowE + lensing + BAO).
Notice that recently, the T2K collaboration~\cite{Abe:2019vii} has published their latest results, which gives the best-fit values of the atmospheric mixing angle $\sin^2 \theta_{23} = 0.53^{+0.03}_{-0.04} $ and the Dirac CP-violating phase $\delta = -1.89^{+0.70}_{-0.58}$ (or $252^{+39.6}_{-32.4}$ in degree) for the normal neutrino mass hierarchy. 
We find for the textures $A_1$  and $A_2$ (see Table~\ref{tab:NuParamNH})   are in well agreement with the latest T2K measurements within the $1\sigma$ confidence level \cite{Abe:2019vii}.

\begin{table}[h]
	\centering  \scriptsize
	\begin{tabular}{|c|c|c|c|} \hline
		Texture & ($m_1, m_2, m_3)\times10^{-2}$ [eV] & $ \sum m_{\nu} $ [eV]  & $ (\delta, \alpha, \beta)^{\circ} $ \\ \hline \hline
   $ A_1(\theta^{bf}_{23}) $  & $ (0.650, 1.067, 5.054)$ & 0.067 & (260, 97, 55)  \\  \hline
   $ A_1(\theta^{max}_{23}) $  & $ (0.564, 1.047, 5.017)$ & 0.066 & (213, 94, 76)  \\  \hline
   $ A_1(\theta^{\rm T2K}_{23}) $  & $ (0.570, 1.067, 4.990)$ & 0.063 & (267, 97, 51)  \\  \hline
   $ A_2(\theta^{bf}_{23}) $  & $(0.466, 0.984, 5.097)$ & 0.065  & (262, 80, 133) \\ \hline
   $ A_2(\theta^{max}_{23}) $  & $(0.577, 1.071, 5.001)$ & 0.066  & (267, 81, 130) \\ \hline
   $ A_2(\theta^{\rm T2K}_{23}) $  & $ (0.504, 1.010, 4.988)$ & 0.065 & (237, 81, 145)  \\  \hline
	\end{tabular}
	\caption{\footnotesize Simulated values for the textures $ A_1, A_2 $ for normal neutrino mass hierarchy. Two sets of solutions are presented for both the textures, which are calculated corresponding to the global best-fit value of $ \theta_{23}$  i.e., $ \theta^{bf}_{23} = 47.7^\circ$~\cite{deSalas:2018bym}, for the maximal value of $ \theta_{23} $ i.e.,  $\theta^{max}_{23}  = 45^\circ$, and for the latest T2K \cite{Abe:2019vii} results, respectively.}
	\label{tab:NuParamNH}
\end{table}

For the textures $ B_3$ and  $ B_4 $, one can have non-zero $ | m_{ee}| $ (see Table \ref{tab:charges}), thus we have predictions for the effective Majorana neutrino mass $| m_{ee}| $  which appears in the neutrinoless double beta ($ 0\nu\beta\beta $) decay experiments. 
At present, the $ 0\nu\beta\beta $ decay $ (A, Z) \longrightarrow (A, Z+2) + 2e^-$ is the unique process which can probe the Majorana nature of massive neutrinos.
%
Currently, number of experiments that are dedicated to look for the signature of $ 0\nu\beta\beta $-decay are namely, GERDA Phase II \cite{Agostini:2018tnm}, CUORE \cite{Alduino:2017ehq}, SuperNEMO \cite{Barabash:2011aa}, KamLAND-Zen \cite{KamLAND-Zen:2016pfg} and EXO \cite{Agostini:2017jim}.
It is to be noted here that, this process violate lepton number by two-units and the half-life of such decay process can be read as \cite{Rodejohann:2011mu,Dev:2013vxa},
\begin{equation}
(T^{0\nu}_{1/2})^{-1} = G_{0\nu}|M_{0\nu}(A,Z)|^{2} | m_{ee}|^{2} \;,
\end{equation}
where $  G_{0\nu}$ is the two-body phase-space factor, and $ M_{0\nu} $ represents the nuclear matrix element (NME). $|m_{ee}|$ is the effective Majorana neutrino mass and  is given by,
\begin{equation}
|m_{ee}| = \left| \sum^3_{i = 1} m_i U^2_{e i} \right| \;,
\end{equation}
where $U$ stands for PMNS mixing matrix as mentioned in Eq.~(\ref{eq:pmns}).

\begin{figure}[h]
	\centering 
	\includegraphics[scale=0.51]{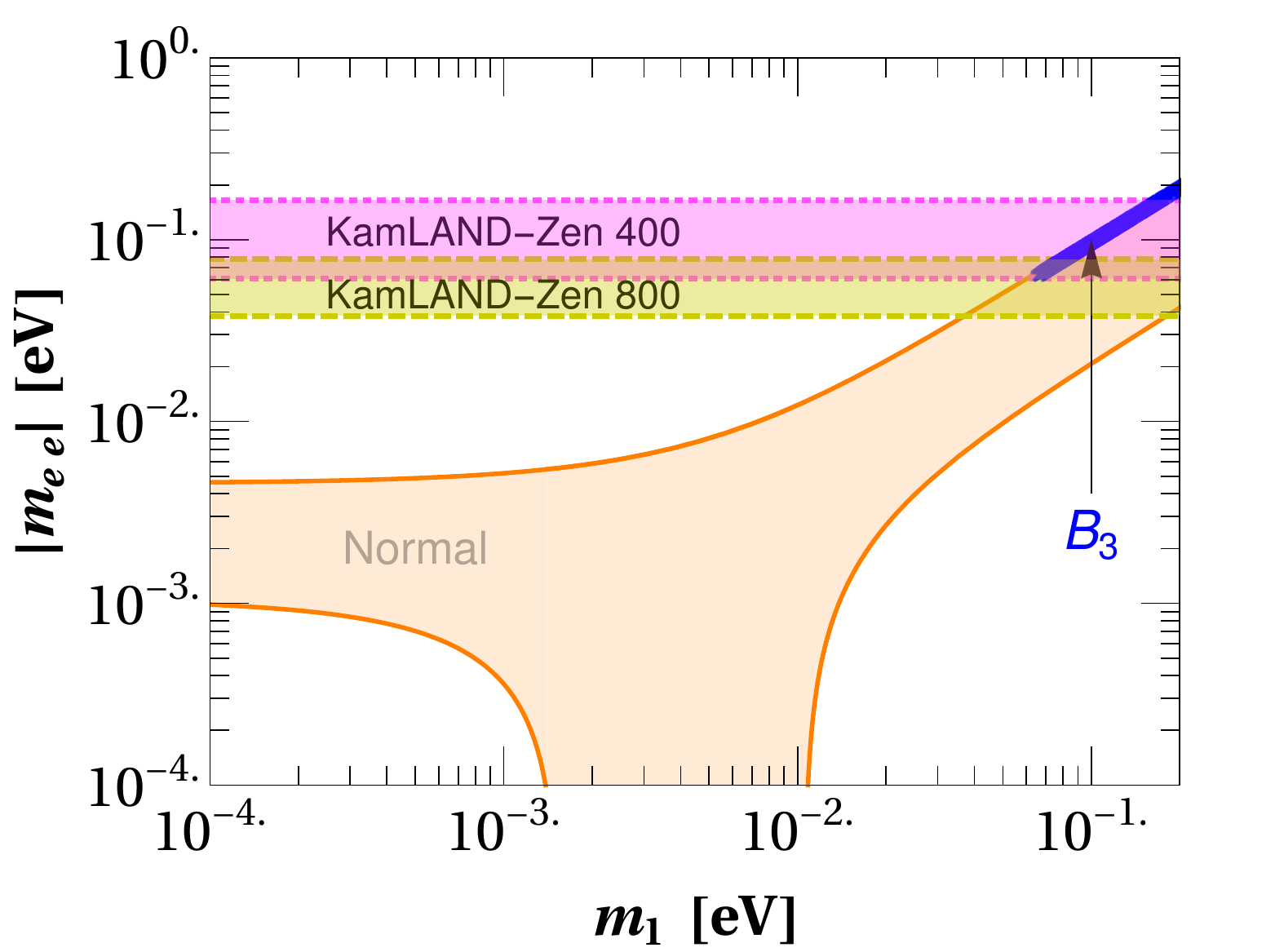}
	\includegraphics[scale=0.51]{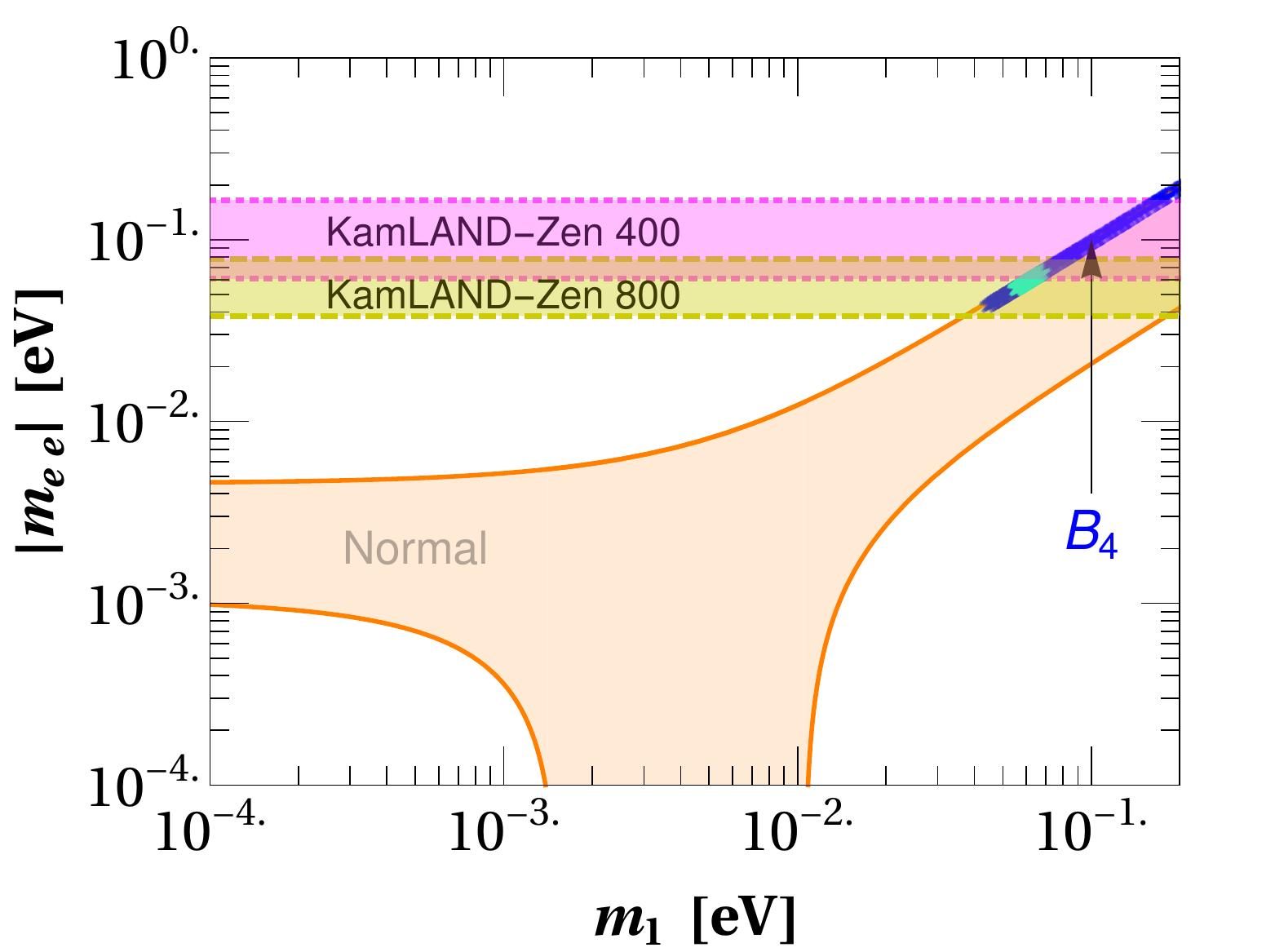}
	\caption{\footnotesize Predictions for the effective Majorana neutrino mass $|m_{ee} |  $ vs the lightest neutrino mass $m_{1}$. The $ 3\sigma $ allowed parameter space of $ |m_{ee}| $ using the latest global-fit data is shown by the light-orange band \cite{deSalas:2018bym}. The bound on $ |m_{ee} |  $ from the KamLAND-Zen 400  \cite{KamLAND-Zen:2016pfg} collaboration has been shown by the light-magenta horizontal band, whereas the first results of the KamLAND-Zen 800 \cite{taup2019} collaboration is outlined by the lighter-green band. Predictions for $|m_{ee} |  $ for $ B_3, $ and  $ B_4 $ are shown by the blue (cyan)  patch at $ 3\sigma $ ($ 1\sigma $). }
	\label{fig:0nu2B}
\end{figure}
%
We present our predictions for the effective Majorana neutrino mass $ |m_{ee}| $ for both the textures in Fig.~\ref{fig:0nu2B}. 
The $ 3\sigma $ allowed parameter space of $ |m_{ee}| $ considering the latest global-fit data \cite{deSalas:2018bym} for the normal neutrino mass hierarchy is shown by the light-orange band \footnote{Note that the present oscillation data tends to favor normal mass hierarchy  (i.e., $\Delta m^2_{31} > 0$) over inverted  mass hierarchy (i.e., $\Delta m^2_{31} < 0$) at more than 3$ \sigma $ \cite{Capozzi:2016rtj,Esteban:2016qun,deSalas:2018bym}, therefor, we focus only on the first scenario.}. 
The magenta band shows the latest bounds on $ |m_{ee}| $, arises from the KamLAND-Zen 400 experiment \cite{KamLAND-Zen:2016pfg} which is read as $  |m_{ee}|  < (61 - 165)$ meV  at 90\% C.L. by taking into account the uncertainty in the estimation of the  nuclear matrix elements.
We also show the first results of KamLAND-Zen 800 collaboration using the lighter-green band, which was presented in the latest meeting TAUP 2019 \cite{taup2019}.
Besides this, the predictions for $ |m_{ee}| $  for the textures $ B_3$ and  $ B_4 $ are shown by the blue (cyan)  patch at $ 3\sigma $ ($ 1\sigma $) significance level. 
We notice from both the panel of Fig.~\ref{fig:0nu2B} that the calculated values of $ |m_{ee}| $ lie in the range $ m_1 \geq 0.06$ eV  for $ B_3$ and $ m_1 \geq 0.04$ eV  for $ B_4 $, respectively.  
It can be seen from the left panel that the predictions of $ B_3$ are in the reach of KamLAND-Zen 400, whereas $ B_4$ predictions can be probed by the KamLAND-Zen 800 data.

It is to be noted here that  the latest bound on the sum of neutrino masses $ \sum m_{\nu} $  come from   \textit{Planck} collaboration~\cite{Aghanim:2018eyx} which gives $ \sum m_{\nu} < 0.12$ eV (95\%, \textit{Planck} TT, TE, EE + lowE + lensing + BAO).  
Now, given the constrained bound on $ \sum m_{\nu} $, if one converts them for the lightest neutrino mass $ m_1 $, then it can be seen that the  textures $ B_3$ is almost rule out. On the other hand, the textures $ B_4$ is consistent with the latest data.
We further examine that none of these textures are able to explain the latest best-fit value of $ \theta_{23}$. 
However, both these types are consistent with the maximal value of the mixing angle $ \theta_{23}$. Considering $ \theta^{max}_{23} $ as a seed point, we calculate remaining unknown in  Table~\ref{tab:NuParamQD} . 
From the fifth column, one can notice that these textures predict maximal value for the Dirac type CP-phase $ \delta $, which is in well agreement with the latest best-fit value within $ 1\sigma $ range \cite{deSalas:2018bym}. Also, CP-conserving values are predicted for the Majorana type CP-phases $ \alpha, \beta $. 
We show the predictions for the  sum of neutrino masses $ \sum m_{\nu} $ and the effective Majorana neutrino mass $ |m_{ee}|$ for texture types $ B_3, $ and  $ B_4 $ in third and fourth column, respectively. 

\begin{table}[h]
	\centering  \scriptsize
	\begin{tabular}{|c|c|c|c|c|} \hline
		Texture & ($m_1 \approx m_2\approx m_3)$ [eV] & $ \sum m $ [eV] & $ \vert\langle m_{ee} \rangle \vert $ [eV] & $ (\delta, \alpha, \beta)^{\circ} $ \\ \hline \hline
   $ B_3(\theta^{max}_{23}) $  & $ 0.144$ & 0.432 & 0.144 & (270, 0, 180)  \\  \hline
   $ B_4(\theta^{max}_{23}) $  & $0.100$ & 0.300  & 0.100 & (270, 180, 0) \\ \hline
	\end{tabular}
	\caption{\footnotesize Simulated values for the textures $ B_3, B_4 $ for quasi-degenerate neutrino mass pattern.}
	\label{tab:NuParamQD}
\end{table}

\section{Conclusion}\label{sec:conclusion}
Physics beyond the Standard Model (BSM), incorporating neutrino masses, are testable in the next generation superbeam neutrino oscillations as well as CE$ \nu $NS experiments.
This work is dedicated to investigating non-standard neutrino interactions (NSIs), a possible sub-leading effects originating from the physics beyond the SM, and eventually can interfere in the measurements of neutrino oscillation parameters.
There exists numbers of BSM scenarios give rise to NSIs that can be
tested in the oscillation experiments. However, such models undergo numerous constrained arising from the different particle physics experiments.
In this work, we focus on an anomaly free $U(1) ^\prime  $ gauge symmetry where a new gauge boson, $ Z^{\prime} $, exchanged has been occurred. Depending on  $U(1) ^\prime  $ charge assignments, we find four different scenarios compatible with the current neutrino oscillation data, namely, $  U(1)_{B-L_\mu-2L_\tau}$, $  U(1)_{B-2L_\mu-L_\tau}$, $  U(1)_{B-L_e-2L_\tau}$, and $  U(1)_{B-L_e-2L_\mu}$.
It has been further realized that these four scenarios correspond to four different two-zero textures for the neutrino mass matrix, namely, $A_1$, $A_2$, $B_3$ and $B_4$. We notice that the NSI parameter 
$ \epsilon_{ee} $ is obtained under  $B_3$ and $B_4$ textures, $A_1$, $A_2$,  and $B_4$ lead to $ \epsilon_{\mu\mu} $, whereas one finds $ \epsilon_{\tau\tau} $ from  $A_1$, $A_2$, and $B_3$.
We summarize our results for possible NSIs considering various experimental limits in Fig. \ref{fig:mass_coupling}, whereas other neutrino phenomenology are given in Fig.  \ref{fig:0nu2B} and in Table \ref{tab:NuParamNH}, \ref{tab:NuParamQD}, respectively.  Depending on our analysis, we make our final remarks as follows:
\begin{itemize}
\item Texture $A_1$:  in this case, we notice that the future COHERENT experiments with NaI or LAr-1t detectors will explore a parameter space for masses  $7 ~{\rm MeV}  \leq M_{Z^\prime} \leq3$ GeV within the coupling limits  $0.8\times10^{-5} \leq g^\prime \leq 10^{-3}$. Also, the parameter space below $5.3$ MeV can be ruled out using the measurement of $ \Delta N_{eff} $ coming from the observation of Big Bang nucleosynthesis. Notice here that this observation holds true for remaining cases. Furthermore, it can be seen that above $3$ GeV. the  LHCb can put the strongest bound. Also, in this scenario, the  effective mass parameter $ m_{ee}$ of the $ 0\nu\beta\beta $-decay is zero.
\item Texture $A_2$: findings of $A_2$ is similar as $A_1$. However, we notice that the future COHERENT experiments will show the tightest constraint upto the mass limit $ \sim 550 $ MeV and above this the LHCb will give the 
stringent bound. It is to be noted here that the LHCb can exclude more parameter space for $A_2$ compared to $A_1$, which is simply because $ \mu-$ field carry 2-units of $U(1) ^\prime  $ charge than of $A_1$ (in case of  $A_1$, $U(1) ^\prime  $ charge of $ \mu-$ field is 1).

 \item Texture $B_3$: outputs of $B_3$ is very different compared to $A_1$ and  $A_2$. Here we notice the CE$ \nu $NS experiment CONUS can explore the most of the parameter space for the masses of $ M_{Z^\prime} $ above $ \sim 25 $ MeV and coupling constant $g^\prime \geq  5\times10^{-6}$. On the other hand, below $25$ MeV, the parameter space has been ruled out by the beam dump experiments.

Moreover, one also have predictions for $ 0\nu\beta\beta $-decay which can be explored by the KamLAND-Zen collaboration (see left panel of Fig. \ref{fig:0nu2B}).
%
\item Texture $B_4$: in this case CONUS can rule out the parameter space  for the  mass range,  $25  \leq M_{Z^\prime} \leq 500$ MeV corresponding to coupling strength $5\times10^{-6} \leq g^\prime \leq 1.5\times10^{-4}$. Above this mass limit and coupling strength the LHCb can put the tightest constraint. Moreover, the beam dump experiments can exclude the parameter space below 25 MeV. 
We also have  predictions for the $ 0\nu\beta\beta $-decay  and the parameter space are marginally consistent with the present limit of both the  KamLAND-Zen and the \textit{Planck} bound as given in the right panel of Fig. \ref{fig:0nu2B}.
\end{itemize}

Finally, we like to emphasize that the $U(1)^{\prime}$ charges that lead to the scenarios $A_1$  and $A_2$, as given in Table~\ref{tab:textures}, the LHCb provides the tightest constraint than the CE$ \nu $NS experiments above 0.55, 3 GeV, respectively.
Moreover, it is noteworthy to notice that the predictions of Dirac CP phase $\delta$ for $A_1$  and $A_2$ (see Table~\ref{tab:NuParamNH}) are in well agreement with the latest T2K result within the $1\sigma$ confidence level \cite{Abe:2019vii}. 
On the other hand, the CE$ \nu $NS experiment CONUS puts the most stringent limit on $B_3$ above $ \sim 25 $ MeV (see the first panel of the second row of Fig. \ref{fig:mass_coupling}). Moreover, the predictions of $B_3$ are in  reach of the KamLAND-Zen 400 data  (see the left panel of the  Fig. \ref{fig:0nu2B}). 
Note further that the $B_4$  is the most constrained one among all the scenarios in the region $25  \leq M_{Z^\prime} \leq 700$ MeV and also the current limits of the $0\nu\beta\beta$ decay coming from the KamLAND-Zen 800 data  are almost excluding this scenario as shown in Figs. \ref{fig:mass_coupling} and \ref{fig:0nu2B}, respectively.
\section{Acknowledgements}
This work is supported by the  German-Mexican  research  collaboration grant SP 778/4-1 (DFG) and 278017 (CONACYT),  CONACYT CB-2017-2018/A1-S-13051 (M\'exico) and DGAPA-PAPIIT IN107118 and  SNI (M\'exico).  NN is supported by the postdoctoral fellowship program DGAPA-UNAM. LJF is supported by a posdoctoral CONACYT grant.

\appendix

\section{Anomaly cancellation conditions}\label{sec:anomalycanc}
For simplicity, let us define $Y' \equiv B-2L_\alpha-L_\beta$ and $U(1)'\equiv U(1)_{Y'}$. The six triangle anomalies of the model are~\cite{Heeck:2012cd}
\begin{subequations}
\begin{align}
	U(1)'-\mathrm{grav}-\mathrm{grav}&: \sum Y' = 9\left( 2(\tfrac{1}{3})-\tfrac{1}{3}-\tfrac{1}{3}\right) + \sum_\ell(2x_\ell - x_\ell)+\sum_i Y'(N_i^c),   \label{eq:anom_a}\\
	U(1)'- U(1)'- U(1)'&: \sum Y'^3 = 9\left( 2(\tfrac{1}{3})^3-(\tfrac{1}{3})^3-(\tfrac{1}{3})^3\right) + \sum_\ell(2x_\ell^3-x_\ell^3) +\sum_i Y'^3(N_i^c), \label{eq:anom_b}\\
	U(1)'- U(1)'- U(1)_Y&: \sum Y'^2Y = 9\left(2(\tfrac{1}{3})^2 (\tfrac{1}{3})+(-\tfrac{1}{3})^2 (-\tfrac{4}{3}) + (-\tfrac{1}{3})^2 (\tfrac{2}{3}) \right) \nonumber \\
	&\phantom{hhhhhhhh}+ \sum_\ell(2x_\ell^2(-1)+x_\ell^2(2)) = 0, \label{eq:anom_c}\\
	U(1)'- U(1)_Y- U(1)_Y&: \sum Y'Y^2 = 9\left(2(\tfrac{1}{3}) (\tfrac{1}{3})^2+(-\tfrac{1}{3}) (-\tfrac{4}{3})^2 + (-\tfrac{1}{3})(\tfrac{2}{3})^2 \right) \nonumber \\
	&\phantom{hhhhhhhh}+\sum_\ell(2x_\ell(-1)^2 - x_\ell(2)^2) = -6 - 2\sum_\ell x_\ell, \label{eq:anom_d}\\
	U(1)'- SU(3)- SU(3)&: \sum_{\bf{3},\bar{3}} Y' = 9\left(2(\tfrac{1}{3})-\tfrac{1}{3}-\tfrac{1}{3} \right) = 0 \label{eq:anom_e}\\
	U(1)'- SU(2)- SU(2)&: \sum_{\bf{2}} Y' = 2(9)(\tfrac{1}{3}) + 2\sum_\ell x_\ell.\label{eq:anom_f} 
\end{align}
\end{subequations}
As it can be seen, conditions in Eq.~\eqref{eq:anom_c} and~\eqref{eq:anom_e} are already equal to zero. By imposing all the other conditions equal to zero, the $U(1)'$ charges of the right-handed neutrinos have to fulfill the following relations
\begin{equation}
    \sum_i Y'(N_i^c) = -\sum_\ell x_\ell = 3, \quad \sum_i Y'^3(N_i^c) = -\sum_\ell x_\ell^3.
\end{equation}
By looking at Table~\ref{tab:charges}, we can notice that these relations hold, since the charges of the right-handed neutrinos are the same as for the charged leptons, and $-\sum_\ell x_\ell= 0+1+2 = 3$.

\section{Neutrino mass matrix}\label{sec:appB}
In this section we will show an example of how to compute the light neutrino mass matrix, for a specific choice of $U(1)'$ charges. Within the type-I seesaw scenario~\cite{Mohapatra:1979ia}, the low energy neutrino mass matrix is given by
\begin{equation}
- m_{\nu} \approx M_D^T M_R^{-1} M_D \;,
\label{eq:lightNeutrinoMassMatrix}
\end{equation}
where $M_D$ and $M_R$ are the Dirac and Majorana neutrino mass matrices, respectively.

In our prescription, the Yukawa Lagrangian invariant under $SM\otimes U(1)^{\prime}$ for the charged-leptons and neutrinos is given by
\begin{eqnarray}
-\mathcal{L}_{Y} & \supset
& y_{e}\overline{L}_{e}\ell_{e}H + y_{\mu}\overline{L}_{\mu}\ell_{\mu} H + y_{\tau}\overline{L}_{\tau} \ell_{\tau} H + y_{1}^{\nu}\overline{L}_{e}\tilde{H}N_1+y_{2}^{\nu}\overline{L}_{\mu}\tilde{H}N_2+y_{3}^{\nu}\overline{L}_{\tau}\tilde{H}N_3 \;.
\end{eqnarray}
This leads to the Dirac neutrino mass matrix of the form
\begin{equation}
M_D = \left( \begin{array}{ccc}
\times & 0 & 0 \\
0 & \times &0 \\
0 & 0 &\times
\end{array}\right)\;.
\end{equation}

For the Majorana neutrino mass matrix, we need to specify the $U(1)'$ fermion charges. For example, with the choice $(x_e, x_\mu, x_\tau) = (0,-1,-2)$, the RH neutrino Lagrangian is
\begin{eqnarray}
-\mathcal{L}_{Majorana} & = 
& \frac{1}{2}M_1\overline{N_1^c} N_1 + \frac{1}{2} y^{N}_1 \overline{N_1^c}N_2\phi_1 + \frac{1}{2} y^{N}_2 \overline{N_1^c}N_3\phi_2+ \frac{1}{2} y^{N}_3 \overline{N_2^c}N_2\phi_2\;.
\end{eqnarray}
Therefore, the Majorana neutrino mass matrix takes the form 
\begin{equation}
M_R = \left( \begin{array}{ccc}
\times & \times & \times \\
\times & \times & 0 \\
\times & 0 & 0
\end{array}\right)\; .
\end{equation}
Plugin $M_D$ and $M_R$ in Eq.~\eqref{eq:lightNeutrinoMassMatrix}, one finds the light neutrino mass matrix of the form
\begin{equation}
m_{\nu} = \left( \begin{array}{ccc}
0 & \times & 0 \\
\times & \times & \times \\
0 & \times & \times
\end{array}\right)\;,
\end{equation}
which corresponds to the type $A_1$ neutrino mass matrix. One can follow the same procedure for the other charge assignments to get the different light neutrino mass matrices (see Table  \ref{tab:textures} for details).

\bibliography{references}

\end{document}